\documentclass[review]{elsarticle}

\usepackage[margin=3cm]{geometry}
\usepackage{epstopdf}
\usepackage[tight,footnotesize,rm,RM]{subfigure}
\usepackage{algorithm}
\usepackage[noend]{algorithmic}
\usepackage{url}
\usepackage{framed}
\usepackage[table]{xcolor}
\usepackage{multirow}
\usepackage{hyperref}
\usepackage{amsthm}
\usepackage{amsfonts,amssymb}
\usepackage{mathtools}
\usepackage{amssymb}
\usepackage{booktabs}
\usepackage{color}
\usepackage[font=footnotesize,skip=0pt]{caption}

\usepackage{natbib}
\newtheorem{mydef}{Definition}[section]

\newtheorem{property}{Property}[section]

\graphicspath{{./figs/}}

\begin{document}

\begin{frontmatter}

\title{Efficient mining of maximal biclusters in mixed-attribute datasets}

\author[myaddress1]{Rosana~Veroneze\corref{mycorrespondingauthor}}
\cortext[mycorrespondingauthor]{Corresponding author. Phone Number: +55(19)99744-3532}
\ead{veroneze@dca.fee.unicamp.br}

\author[myaddress1]{Fernando~J.~Von~Zuben}

\address[myaddress1]{University of Campinas (DCA/FEEC), 400 Albert Einstein Street, Campinas, SP, Brazil – 13083-852}

\begin{abstract}
This paper presents a novel enumerative biclustering algorithm to directly mine all maximal biclusters in mixed-attribute datasets, with or without missing values. The independent attributes are mixed or heterogeneous, in the sense that both numerical (real or integer values) and categorical (ordinal or nominal values) attribute types may appear together in the same dataset. The proposal is an extension of RIn-Close\_CVC, which was originally conceived to mine perfect or perturbed biclusters with constant values on columns solely from numerical datasets, and without missing values. Even endowed with additional and more general features, the extended RIn-Close\_CVC retains four key properties: (1) efficiency (take polynomial time per bicluster), (2) completeness (find all maximal biclusters), (3) correctness (all biclusters attend the user-defined measure of internal consistency), and (4) non-redundancy (all the obtained biclusters are maximal and the same bicluster is not enumerated twice). Our proposal is the first one to deal with mixed-attribute datasets without requiring any pre-processing step, such as discretization and itemization of real-valued attributes. This is a decisive aspect, because discretization and itemization implies a priori decisions, with information loss and no clear control over the consequences. On the other hand, even having to specify a priori an individual threshold for each numerical attribute, that will be used to indicate internal consistency per attribute, each threshold will be applied during the construction of the biclusters, shaping the peculiarities of the data distribution. We also explore the strong connection between biclustering and frequent pattern mining to (1) provide filters to select a compact bicluster set that exhibits high relevance and low redundancy, and (2) in the case of labeled datasets, automatically present the biclusters in a user-friendly and intuitive form, by means of quantitative class association rules. Our experimental results showed that the biclusters yield a parsimonious set of relevant rules, providing useful and interpretable models for five mixed-attribute labeled datasets.
\end{abstract}


\end{frontmatter}

\section{Introduction}
\label{sec:intro}

Given a data matrix composed of a set of objects in its rows and their corresponding attributes in its columns, biclustering is a data mining technique characterized by the simultaneous clustering of both rows and columns of the data matrix, aiming at revealing highly consistent patterns in sub-matrices \cite{ChengChurch2000}. The biclustering result cannot be achieved by two sequential clustering steps, and the internal consistency of each bicluster may involve more general affinity measures than those usually associated with conventional clustering approaches. In fact, a bicluster may be interpreted as a local model, clearly indicating which subset of attributes is responsible for keeping those objects together. As a consequence, when more flexible biclustering structures are considered, such as the ones admitting arbitrarily positioned and overlapping biclusters, as will be the case in this paper, any object or attribute of the data matrix may belong to none, one, or more than one of the obtained biclusters \cite{MadeiraOliveira2004}.

Besides those distinctive aspects when compared to conventional clustering, the biclustering problem has a strong connection with several other relevant problems in multivariate data analysis, including subspace clustering \cite{KriegelEtAl2009}, formal concept analysis (FCA) \cite{Ganter1997}, frequent pattern mining (FPM) \cite{ceglarEtAl2006, HippEtAl2000}, and graph theory. In the subspace clustering area, the biclustering problem is called \emph{pattern-based clustering} \cite{KriegelEtAl2009}. The problem of mining the \emph{concept lattice} (i.e., to enumerate all \emph{formal concepts}) from a \emph{formal context} is the same as enumerating all maximal biclusters of ones from a binary data matrix \cite{Kaytoue2011}, and it is the same as enumerating all \emph{maximal bicliques} from a \emph{bipartite graph}. Besides that, the \emph{intent} of a formal concept is the same as a \emph{closed itemset} \cite{LakhalStumme2005}. Several algorithms of FCA and FPM, which are restricted to binary datasets, such as In-Close2 \cite{Andrews2011} and Charm \cite{ZakiEtAL2002}, are characterized by exhibiting four key properties: (1) efficiency (take polynomial time per bicluster), (2) completeness (find all maximal biclusters), (3) correctness (all biclusters attend the user-defined measure of similarity), and (4) non-redundancy (all the obtained biclusters are maximal and the same bicluster is not enumerated twice). So, very powerful biclustering algorithms have already been proposed to deal with binary datasets.

Recently, Veroneze \emph{et al.} \cite{VeronezeEtAl2017} proposed a family of algorithms, called RIn-Close, also exhibiting those four key properties when enumerating biclusters directly in numerical (not only binary, but also integer or real-valued) data matrices. It may be considered a significant achievement, given that, before the RIn-Close family of algorithms, finding biclusters in numerical data matrices was accomplished by algorithms not exhibiting those four properties, or by discretizing and itemizing the numerical matrix, ultimately treating binary matrices, which implies information loss \cite{Besson2007,SrikantAgrawal1996}. Notice that the RIn-Close family of algorithms is capable of mining perfect and perturbed biclusters with constant values on rows (CVR) and constant values on columns (CVC), and also perfect biclusters with coherent values (CHV). There is also an algorithm to enumerate perturbed CHV biclusters, but in this case the algorithm can not be considered efficient, due to the necessity of dealing with expanded matrices and mining the CHV biclusters from CVC biclusters \cite{VeronezeEtAl2017}.

Motivated by the existence of relevant practical biclustering problems in which numerical (discrete or continuous) and categorical (ordinal or nominal) attributes are simultaneously present in the same dataset, we are going to propose here and extension of one of the RIn-Close algorithms to directly treat those kind of mixed-attribute datasets, retaining the four key properties and thus enlarging the applicability of the RIn-Close family. The authors are aware of the existence in the literature of alternative biclustering proposals to deal with mixed-attribute datasets (see Section~\ref{sec:relwork} for more details), but none of those existing proposals exhibits the four key properties or, when the four key properties are present, the numerical attributes should pass through discretization and itemization before the mining process, which inevitably promote information loss. So, we are going to present in this paper the first enumerative biclustering algorithm with those four key properties to directly mine all maximal biclusters in a mixed-attribute dataset, without the necessity of any pre-processing step. It is worth anticipating that ($i$) according to the convenience of the user, pre-processing, such as normalization, scaling, or discretization of any attribute, is fully admissible, thus being optional, but not mandatory; ($ii$) fully numerical or even fully categorical datasets are special cases of mixed-attribute datasets, being promptly treatable by our new proposal as well.

In Section~\ref{sec:bic}, we will formally present the types of biclusters that can be mined from a data matrix. Most importantly, we will demonstrate that CVC biclusters are the only type of biclusters that makes an immediate sense when mixed-attribute datasets are considered. Therefore, we firstly extend the definition of a CVC bicluster provided by \cite{VeronezeEtAl2017}, so that it will work with numerical and/or categorical attributes. Subsequently, we will generalize RIn-Close\_CVC \cite{VeronezeEtAl2017} to enumerate all maximal CVC biclusters in mixed-attribute datasets. Even when handling a strictly numerical data matrix, the previous version of RIn-Close\_CVC requires normalization or scaling of the real-valued attributes, particularly when the range of the attributes are very different. The extended version of RIn-Close\_CVC, to be proposed here, makes this pre-processing step optional, in the sense that the final results will not be influenced by normalizing or scaling any attribute (column of the data matrix), supposing the user has properly defined the consistency threshold for each numerical column. So, we also have a threshold to decide if a row or a column will enter a given bicluster, but here the threshold will be applied over the original attribute values, without information loss.

An additional advantage of the extended version of RIn-Close\_CVC is the ability to directly handle missing values, without the necessity of performing an imputation step, as required by the original version of RIn-Close\_CVC \cite{VeronezeEtAl2017}.

Essentially, we are going to propose a general-purpose and low-cost enumerative algorithm devoted to biclustering mixed-attribute datasets, characterized by no information loss and no introduction of additional noise. Besides, the sparser the matrix, the faster the enumeration tends to be \cite{Veroneze2016}. This is because our proposal simply ignores the missing elements of the data matrix. Notice that this ability to deal with missing data can be easily incorporated to the other RIn-Close algorithms, as was done by Veroneze \cite{Veroneze2016}.

Given that enumerative biclustering algorithms may return a huge amount of biclusters, part of them being useless or at least of low relevance, the contribution of this paper goes further. We have explored the strong connections between biclustering and FPM, borrowing metrics generally adopted to evaluate association rules (AR) \cite{AgrawalEtAl1993} and class association rules (CAR) \cite{LiuEtAl1998} to select a subset of relevant biclusters, where relevance may be associated with user-defined thresholds for these FPM metrics. The motivation for not adopting other internal evaluation metrics available in the literature, such as the \emph{Mean Squared Error} \cite{ChengChurch2000}, is the lack of consensus about which is the most indicated, particularly in the context of mixed-attribute datasets. We also have several proposals for external evaluation metrics \cite{HortaCampello2014}, but they require a reference solution, not attainable in real datasets.

Nonetheless, even using a filter based on FPM metrics, the number of biclusters may still be too much for a manual inspection of each bicluster. So, we will incorporate a simple greedy heuristic to select an even smaller subset of the enumerated biclusters to present to the user. This twice-reduced subset of biclusters keeps the same object coverage when compared to the output of the aforementioned FPM relevance filter, aiming at preserving representativeness. These two filters are similar to the ones proposed by Veroneze \cite{Veroneze2016}. In that work, RIn-Close was adapted to mine only \emph{pure biclusters} from a labeled dataset, which are full confident biclusters, in the sense of being composed of objects sharing the same class label. Also, the greedy heuristic presented here prioritizes the biclusters with small intents, being a slightly different version of the one proposed in \cite{Veroneze2016}.

Based on the results provided by this cascade of two filters, we will discuss the potential of the biclustering approach to provide interpretative models for labeled datasets. In fact, it is not easy for the user to properly interpret a biclustering solution. We argue that \emph{quantitative association rules} (QARs) \cite{zhu2009} and \emph{quantitative class association rules} (QCARs) are simple and interpretative formats to present biclusters to the user. For instance, using QCARs directly extracted from the biclusters, the user is informed about the attributes involved, their range of values and the associated class that is being represented.

The remainder of the paper is organized as follows. Section~\ref{sec:bic} introduces definitions and mathematical notation for biclustering, and also specifically for mixed-attribute biclustering. Section~\ref{sec:CAR} reviews some FPM definitions and metrics, and describes the two filters used in this work to select biclusters. Section~\ref{sec:relwork} is devoted to related works. The extended version of RIn-Close\_CVC is presented in Section~\ref{sec:rinclose}. Experimental results are discussed in Section~\ref{sec:exp}. Concluding remarks and further steps of the research are outlined in Section~\ref{sec:conclusion}.

\section{Biclustering}
\label{sec:bic}

The formalism used here to describe a bicluster and its variants is based on \cite{VeronezeEtAl2017}.

Let $\mathbf{A}_{n \times m}$ be a data matrix with the row index set $X = \left \{ 1, 2,..., n \right \}$ and the column index set $Y = \left \{ 1, 2, ...,m \right \}$. Each row represents an object, and each column represents an attribute. Each element $a_{ij} \in \mathbf{A}$ represents the relationship between object $i$ and attribute $j$. We use $(X,Y)$ to denote the entire matrix $\mathbf{A}$. Considering that $I \subseteq X$ and $J \subseteq Y$, $\mathbf{A}_{IJ} = (I, J)$ denotes the submatrix of $\mathbf{A}$ with the row index subset $I$ (named \emph{extent} in FCA) and column index subset $J$ (named \emph{intent} in FCA).

\begin{mydef}
A bicluster is a submatrix $(I,J)$ of the data matrix $\mathbf{A}_{n \times m}$ such that the rows in the index subset $I = \left \{ i_1,..., i_k \right \}$ ($I \subseteq X$ and $k \leq n$) exhibit a consistent pattern across the columns in the index subset $J = \left \{ j_1,..., j_s \right \}$ ($J \subseteq Y$ and $s \leq m$), and vice-versa.
\label{def:bic}
\end{mydef}

Thus, a bicluster $(I,J)$ is a $k \times s$ submatrix of the matrix $\mathbf{A}$, not necessarily with contiguous rows and columns, such that it meets a certain homogeneity criterion. A biclustering algorithm looks for a set of biclusters $\mathfrak{B} = (I_l, J_l)$, $l = 1, ..., q$, such that each bicluster $(I_l, J_l)$ satisfies some specific characteristics of homogeneity \cite{MadeiraOliveira2004}. Considering these characteristics, there are four major types of biclusters \cite{MadeiraOliveira2004}: ($i$) biclusters with constant values (CTV), ($ii$) biclusters with constant values on columns (CVC) or rows (CVR), ($iii$) biclusters with coherent values (CHV), and ($iv$) biclusters with coherent evolutions (CHE). There are many subtypes of CHE biclusters, and the order-preserving submatrix (OPSM) biclusters are the most famous among them. The total number of biclusters, $q$, will depend on the features of the selected biclustering algorithm, on the constraints imposed, and on the behaviour of the dataset being analysed.

\subsection{Types of Biclusters}
\label{sec:bicTypes}

\begin{figure*}
  \centering
	\subfigure[Perfect biclusters.]{
		\includegraphics[trim=2.5cm 13cm 7.5cm 2cm, clip, scale=0.65]{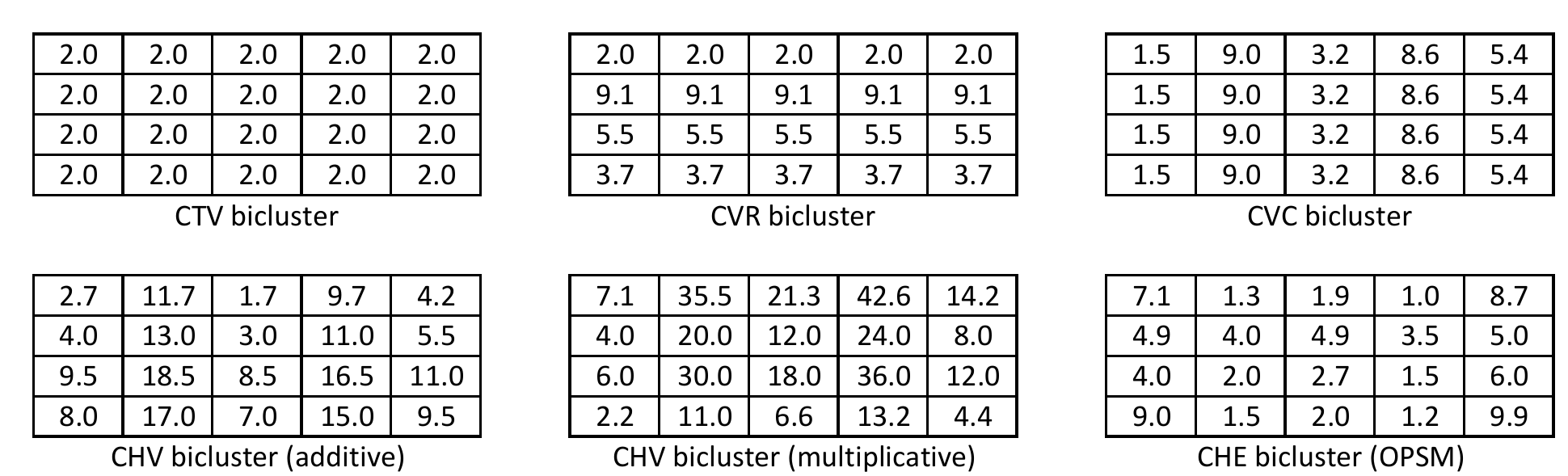}
		\label{fig:typesOfBica}
	}
	\subfigure[Perturbed biclusters.]{
		\includegraphics[trim=2.5cm 13cm 7.4cm 2cm, clip, scale=0.65]{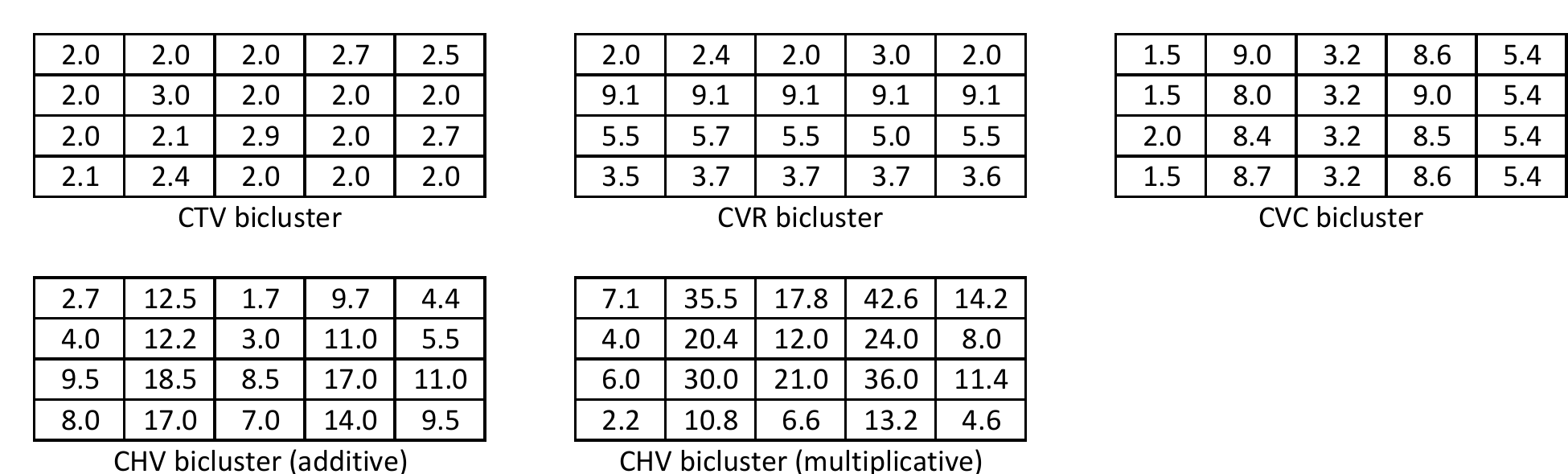}
		\label{fig:typesOfBicb}
	}
  \caption{Examples of different types of biclusters (extracted from \cite{VeronezeEtAl2017}).}
  \label{fig:typesOfBic}
\end{figure*}

Although perfect biclusters can be found in some data matrices, they are usually masked by noise in real datasets. Therefore, a user-defined parameter $\epsilon \geq 0$ determines the maximum residue (perturbation) allowed in a bicluster. Perfect biclusters are mined using $\epsilon = 0$, whereas perturbed biclusters are mined using $\epsilon > 0$. Dealing with numerical data matrices, the RIn-Close family of algorithms has specialized algorithms for mining perfect biclusters that are faster than the algorithms for mining perturbed biclusters. Fig.~\ref{fig:typesOfBic} shows examples of different types of numerical biclusters, in both perfect and perturbed cases.

\begin{mydef}[CTV biclusters]
A \emph{CTV bicluster} is a submatrix $(I, J)$ of a data matrix $\mathbf{A}_{n \times m}$ such that

\begin{equation}
  \max_{i \in I, j \in J} (a_{ij}) - \min_{i \in I, j \in J} (a_{ij}) \leq \epsilon.
	\label{eq:ctvbic}
\end{equation}

\label{def:ctvbic}
\end{mydef}

Fig.~\ref{fig:typesOfBica} shows an example of a perfect CTV bicluster, and Fig.~\ref{fig:typesOfBicb} shows an example of a perturbed CTV bicluster that can be mined using $\epsilon \geq 1$.

\begin{mydef}[CVC biclusters]
A \emph{CVC bicluster} is a submatrix $(I, J)$ such that

\begin{equation}
  \max_{i \in I} (a_{ij}) - \min_{i \in I} (a_{ij}) \leq \epsilon, \forall j \in J.
	\label{eq:cvcbic}
\end{equation}

\label{def:cvcbic}
\end{mydef}

Fig.~\ref{fig:typesOfBica} shows an example of a perfect CVC bicluster, and Fig.~\ref{fig:typesOfBicb} shows an example of a perturbed CVC bicluster that can be mined using $\epsilon \geq 1$.

The definition of a CVR bicluster is the equivalent transpose of the definition of a CVC bicluster. So, we can mine CVR biclusters by transposing the original data matrix and using an algorithm to mine CVC biclusters. See examples of CVR biclusters in Fig.~\ref{fig:typesOfBic}.

There are two perspectives for CHV biclusters: (\textit{i}) additive model, and (\textit{ii}) multiplicative model (see Fig.~\ref{fig:typesOfBic}). Biclusters based on the additive model are called \textit{shifting biclusters}. Biclusters based on the multiplicative model are called \textit{scaling biclusters}. Any row (column) of a perfect shifting bicluster can be obtained by adding a constant value to any other row (column) of the bicluster. Similarly, any row (column) of a perfect scaling bicluster can be obtained by multiplying a constant value to any other row (column) of the bicluster. The problems of mining shifting and scaling biclusters are equivalent. Using an algorithm to mine shifting (scaling) biclusters, we can mine scaling (shifting) biclusters by previously taking the logarithm (exponential) of all entries of the data matrix. Therefore, we are going to provide only the definition of a shifting bicluster in this paper.

\begin{mydef}[CHV biclusters - additive model]
Let $Z^{jl} = \{a_{ij} - a_{il}\}_{i \in I}$, $j,l \in J$, be the set of values of the difference between two attributes for the subset of rows $I$. A \emph{shifting bicluster} is a submatrix $(I,J)$ such that

\begin{equation}
  \max(Z^{jl}) - \min(Z^{jl}) \leq \epsilon, \forall j,l \in J.
	\label{eq:chvabic}
\end{equation}

\label{def:chvabic}
\end{mydef}

Algorithms for finding CHE (coherent evolution) biclusters address the problem of finding coherent evolutions across the rows and/or columns of the data matrix, regardless of their exact values \cite{MadeiraOliveira2004}. The OPSM biclusters are the most famous among the CHE biclusters.

\begin{mydef}[OPSM biclusters]
An \emph{OPSM bicluster} is a submatrix $(I, J)$ of a data matrix $\mathbf{A}_{n \times m}$ such that there is a permutation $P = \{p_1, p_2, ..., p_s\}$ of the set of columns $J$, where $a_{ip_1} \leq a_{ip_2} \leq ... \leq a_{ip_s}$, $\forall i\in I$.

\label{def:opsmbic}
\end{mydef}

Figure~\ref{fig:typesOfBic} shows an OPSM bicluster, where $P = \{4, 2, 3, 1, 5\}$.

\subsection{Maximality and Algebraic Properties}
\label{sec:bicMaxProp}

\begin{mydef}[Maximal bicluster]
Given the desired characteristics of homogeneity, a bicluster $(I,J)$ is called a \emph{maximal bicluster} if and only if:
\begin{itemize}
	\item $\forall x \in X \setminus I$, $(I \cup \{x\}, J)$ is not a (valid) bicluster, and
	\item $\forall y \in Y \setminus J$, $(I, J \cup \{y\})$ is not a (valid) bicluster.
\end{itemize}
\end{mydef}

\noindent It means that a bicluster is maximal if we cannot add any object/attribute without violating the desired characteristics of homogeneity. For instance, a CTV bicluster $(I,J)$ is called a maximal CTV bicluster iff:
\begin{itemize}
	\item $\forall x \in X \setminus I$, $\max_{i \in I \cup \{x\}, j \in J} (a_{ij}) - \min_{i \in I \cup \{x\}, j \in J} (a_{ij}) > \epsilon$, and
	\item $\forall y \in Y \setminus J$, $\max_{i \in I, j \in J \cup \{y\}} (a_{ij}) - \min_{i \in I, j \in J \cup \{y\}} (a_{ij}) > \epsilon$.
\end{itemize}

For all bicluster definitions in Subsection~\ref{sec:bicTypes}, we have the following algebraic properties.

\begin{property}[Anti-Monotonicity]
Let $(I,J)$ be a bicluster. Any submatrix $(I', J')$, where $I' \subseteq I$ and $J' \subseteq J$, is also a (valid) bicluster.
\end{property}

\begin{property}[Monotonicity]
Let $(I,J)$ be a maximal bicluster. Any supermatrix of $(I,J)$ is not a (valid) bicluster.
\end{property}

Usually, the efficient enumerative algorithms of FCA and FPM areas are based on the monotonicity and anti-monotonicity properties \cite{Besson2007}. In fact, we do not know any FCA / FPM efficient algorithm that is not based on these properties. Notice that the RIn-Close family of algorithms also holds both properties.

\subsection{Mixed-Attribute Biclustering}

A dataset may have \emph{numerical} and \emph{categorical} attributes. The numerical attributes can be \emph{discrete} (integer attributes) or \emph{continuous} (real-valued attributes). The categorical attributes can be \emph{ordinal} (attributes that have some kind of implicit or natural order) and \emph{nominal} (attributes that do not have an implicit or natural order or rank). Binary attributes can be seen as nominal attributes that can assume only two values, such as \emph{Yes} or \emph{No}.

\begin{mydef}
A \emph{mixed-attribute dataset} is a dataset that contains more than one attribute type associated with its columns.
\label{def:mixedData}
\end{mydef}

Table~\ref{tab:maDataEx1} shows an example of a mixed-attribute dataset. The attributes \emph{Sex}, \emph{Smoker}, and \emph{Religion} are nominal attributes, with \emph{Sex} and \emph{Smoker} being binary attributes. \emph{Social Class} is an ordinal attribute, where the label \emph{E} represents the poorest people, and the label \emph{A} represents the richest people. \emph{Age} is an integer attribute. \emph{Weight} and \emph{Height} are real-valued attributes.

\linespread{1}
\begin{table}[]
\footnotesize
\centering
\caption{Example of a mixed-attribute dataset, with two perturbed biclusters highlighted. There are 20 objects and 7 attributes.}
\label{tab:maDataEx1}
\begin{tabular}{cccrrclc}
\toprule
\textbf{\#} & \textbf{Sex} & \textbf{Age} & \textbf{Weight (kg)} & \textbf{Height (m)} & \textbf{Smoker} & \textbf{Religion} & \textbf{Social Class} \\
\midrule
1   & \colorbox[rgb]{0.7,0.7,0.7}{F}            & 32           & 94.87                & 1.72                & Y               & \colorbox[rgb]{0.7,0.7,0.7}{Christian}         & \colorbox[rgb]{0.7,0.7,0.7}{C}                     \\
2           & F            & 34           & 99.39                & 1.63                & N               & Christian         & D                     \\
3           & F            & 33           & 124.15               & 1.66                & N               & Hindu             & C                     \\
4           & M            & 52           & 49.77                & 1.71                & Y               & Christian         & E                     \\
5           & F            & 57           & 65.13                & 1.80                & N               & Hindu             & C                     \\
6           & F            & 39           & 58.71                & 1.74                & N               & Buddhist          & E                     \\
7  & \colorbox[rgb]{0.7,0.7,0.7}{F}            & 39           & 67.41                & 1.56                & N               & \colorbox[rgb]{0.7,0.7,0.7}{Christian}         & \colorbox[rgb]{0.7,0.7,0.7}{C}                     \\
8  & \colorbox[rgb]{0.7,0.7,0.7}{F}            & 47           & 67.19                & 1.79                & Y               & \colorbox[rgb]{0.7,0.7,0.7}{Christian}         & \colorbox[rgb]{0.7,0.7,0.7}{B}                     \\
9           & M            & 58           & 42.95                & 1.48                & N               & Christian         & A                     \\
10  & \colorbox[rgb]{0.9,0.9,0.9}{M}            & 17           & 109.52               & \colorbox[rgb]{0.9,0.9,0.9}{1.62}                & \colorbox[rgb]{0.9,0.9,0.9}{N}               & Christian         & \colorbox[rgb]{0.9,0.9,0.9}{C}                     \\
11          & F            & 42           & 91.12                & 1.76                & N               & Buddhist          & D                     \\
12          & F            & 48           & 58.07                & 1.50                & N               & Islamist          & D                     \\
13  & \colorbox[rgb]{0.9,0.9,0.9}{M}            & 43           & 46.69                & \colorbox[rgb]{0.9,0.9,0.9}{1.61}                & \colorbox[rgb]{0.9,0.9,0.9}{N}               & Hindu             & \colorbox[rgb]{0.9,0.9,0.9}{B}                     \\
14  & \colorbox[rgb]{0.9,0.9,0.9}{M}            & 55           & 85.38                & \colorbox[rgb]{0.9,0.9,0.9}{1.54}                & \colorbox[rgb]{0.9,0.9,0.9}{N}               & Islamist          & \colorbox[rgb]{0.9,0.9,0.9}{C}                     \\
15          & M            & 34           & 39.77                & 1.70                & N               & Christian         & B                     \\
16          & M            & 34           & 83.90                & 1.74                & N               & Islamist          & D                     \\
17          & M            & 51           & 55.72                & 1.93                & Y               & Islamist          & B                     \\
18  & \colorbox[rgb]{0.7,0.7,0.7}{F}            & 47           & 57.10                & 1.51                & N               & \colorbox[rgb]{0.7,0.7,0.7}{Christian}         & \colorbox[rgb]{0.7,0.7,0.7}{C}                     \\
19          & M            & 38           & 54.01                & 1.85                & Y               & Islamist          & C                     \\
20  & \colorbox[rgb]{0.9,0.9,0.9}{M}            & 45           & 73.10                & \colorbox[rgb]{0.9,0.9,0.9}{1.59}                & \colorbox[rgb]{0.9,0.9,0.9}{N}               & Islamist          & \colorbox[rgb]{0.9,0.9,0.9}{C}                     \\
\bottomrule
\end{tabular}
\end{table}
\linespread{1.5}

Given the bicluster types presented in Subsection~\ref{sec:bicTypes}, we argue that only CVC biclusters make a direct sense in mixed-attribute datasets - see, for instance, the two highlighted biclusters in Table~\ref{tab:maDataEx1}. Clearly, CVR, CTV, CHV and OPSM biclusters can not be properly characterized with heterogeneous attributes, because attributes of a distinct nature can not be directly related to each other, as required by those bicluster types.

However, Definition~\ref{def:cvcbic} that was already provided to describe CVC biclusters is specific for numerical datasets. Moreover, Definition~\ref{def:cvcbic} also considers that all attributes assume values in the same range, because the same value of $\epsilon$ is adopted for all attributes, which requires a normalization pre-processing step. Therefore, to account for mixed-attribute datasets, the definition of CVC biclusters must be generalized accordingly.

Notice that categorical attributes are discrete entities. The domain of a discrete attribute can be represented by a set of symbols. In the ordinal case, a set of integer values obeying a bijective mapping is a straightforward choice. Thus, we can use integer attributes instead of ordinal attributes without any loss of information. In the nominal case, a one-hot binary representation is generally adopted to impose the same Hamming distance between any pair of distinct values of that attribute. However, in the case of two categories, a single bit is enough. For instance, the values of the nominal attributes \emph{Sex} and \emph{Smoker} may be represented by a bit of information, while the nominal attributes associated with \emph{Religion} may be mapped to one-hot binary sequences. The values \emph{F} and \emph{M} of the \emph{Sex} attribute are mapped to 0 and 1, respectively. The values \emph{N} and \emph{Y} of the \emph{Smoker} attribute are mapped to 0 and 1, respectively. The values \emph{Christian}, \emph{Islamist}, \emph{Hindu}, and \emph{Buddhist} of the \emph{Religion} attribute are mapped to 1000, 0100, 0010 and 0001, respectively. On the other hand, the discrete values of the ordinal attribute \emph{Social Class}, \emph{A}, \emph{B}, \emph{C}, \emph{D}, and \emph{E}, are mapped to 1, 2, 3, 4 and 5, respectively.

Before generalizing the definition of a CVC bicluster, it is possible to simplify even more the numerical representation of a nominal attribute. Given that a nominal attribute is discrete and finite, and supposing we are just focusing on detecting if the attribute value is equal or not, then we may convert each nominal attribute to a distinct integer, producing a more concise numerical representation than one-hot binary representation. We are aware that this representation imposes an arbitrary ordinal relation among the previous nominal attribute values, but this ordinal relation will not affect the results, being an internal manipulation transparent to the user. So, the values \emph{Christian}, \emph{Islamist}, \emph{Hindu}, and \emph{Buddhist} of the \emph{Religion} attribute can be mapped to 1, 2, 3 and 4, respectively. Table~\ref{tab:maDataEx2} contains only numbers and has essentially the same information of Table~\ref{tab:maDataEx1}.

\linespread{1}
\begin{table}[]
\footnotesize
\centering
\caption{Mixed-attribute dataset of Table~\ref{tab:maDataEx1} with its categorical attributes mapped to integer and binary representations. The two highlighted biclusters here are equivalent to the two highlighted biclusters in Table~\ref{tab:maDataEx1}.}
\label{tab:maDataEx2}
\begin{tabular}{cccrrclc}
\toprule
\textbf{\#} & \textbf{Sex} & \textbf{Age} & \textbf{Weight (kg)} & \textbf{Height (m)} & \textbf{Smoker} & \textbf{Religion} & \textbf{Social Class} \\
\midrule
1  & \colorbox[rgb]{0.7,0.7,0.7}{0}   & 32  & 94.87       & 1.72       & 1      & \colorbox[rgb]{0.7,0.7,0.7}{1}        & \colorbox[rgb]{0.7,0.7,0.7}{3}  \\
2  & 0   & 34  & 99.39       & 1.63       & 0      & 1        & 4  \\
3  & 0   & 33  & 124.15      & 1.66       & 0      & 3        & 3  \\
4  & 1   & 52  & 49.77       & 1.71       & 1      & 1        & 5  \\
5  & 0   & 57  & 65.13       & 1.80       & 0      & 3        & 3  \\
6  & 0   & 39  & 58.71       & 1.74       & 0      & 4        & 5  \\
7  & \colorbox[rgb]{0.7,0.7,0.7}{0}   & 39  & 67.41       & 1.56       & 0      & \colorbox[rgb]{0.7,0.7,0.7}{1}        & \colorbox[rgb]{0.7,0.7,0.7}{3}  \\
8  & \colorbox[rgb]{0.7,0.7,0.7}{0}   & 47  & 67.19       & 1.79       & 1      & \colorbox[rgb]{0.7,0.7,0.7}{1}        & \colorbox[rgb]{0.7,0.7,0.7}{2}  \\
9  & 1   & 58  & 42.95       & 1.48       & 0      & 1        & 1  \\
10 & \colorbox[rgb]{0.9,0.9,0.9}{1}   & 17  & 109.52      & \colorbox[rgb]{0.9,0.9,0.9}{1.62}       & \colorbox[rgb]{0.9,0.9,0.9}{0}      & 1        & \colorbox[rgb]{0.9,0.9,0.9}{3}  \\
11 & 0   & 42  & 91.12       & 1.76       & 0      & 4        & 4  \\
12 & 0   & 48  & 58.07       & 1.50       & 0      & 2        & 4  \\
13 & \colorbox[rgb]{0.9,0.9,0.9}{1}   & 43  & 46.69       & \colorbox[rgb]{0.9,0.9,0.9}{1.61}       & \colorbox[rgb]{0.9,0.9,0.9}{0}      & 3        & \colorbox[rgb]{0.9,0.9,0.9}{2}  \\
14 & \colorbox[rgb]{0.9,0.9,0.9}{1}   & 55  & 85.38       & \colorbox[rgb]{0.9,0.9,0.9}{1.54}       & \colorbox[rgb]{0.9,0.9,0.9}{0}      & 2        & \colorbox[rgb]{0.9,0.9,0.9}{3}  \\
15 & 1   & 34  & 39.77       & 1.70       & 0      & 1        & 2  \\
16 & 1   & 34  & 83.90       & 1.74       & 0      & 2        & 4  \\
17 & 1   & 51  & 55.72       & 1.93       & 1      & 2        & 2  \\
18 & \colorbox[rgb]{0.7,0.7,0.7}{0}   & 47  & 57.10       & 1.51       & 0      & \colorbox[rgb]{0.7,0.7,0.7}{1}        & \colorbox[rgb]{0.7,0.7,0.7}{3}  \\
19 & 1   & 38  & 54.01       & 1.85       & 1      & 2        & 3  \\
20 & \colorbox[rgb]{0.9,0.9,0.9}{1}   & 45  & 73.10       & \colorbox[rgb]{0.9,0.9,0.9}{1.59}       & \colorbox[rgb]{0.9,0.9,0.9}{0}      & 2        & \colorbox[rgb]{0.9,0.9,0.9}{3}  \\
\bottomrule
\end{tabular}
\end{table}
\linespread{1.5}

Now we are ready to propose a simple generalization of Definition~\ref{def:cvcbic}, associated with CVC biclusters, to allow an immediate manipulation of mixed-attribute datasets. We are going to use one particular $\epsilon$ per column, and every time that a nominal attribute is being manipulated in a specific column of the mixed-attribute dataset, $\epsilon$ should be taken as zero. On the other hand, for categorical attributes exhibiting an ordinal relation, a suitable integer value should be adopted for $\epsilon$ (it will depend on what the user wants to accept as being part of the same group).

\begin{mydef}[CVC biclusters]
A \emph{CVC bicluster} is a submatrix $(I, J)$ such that

\begin{equation}
  \max_{i \in I} (a_{ij}) - \min_{i \in I} (a_{ij}) \leq \epsilon_j, \forall j \in J,
	\label{eq:cvcbic2}
\end{equation}

\noindent where $\epsilon_j$ is the user-defined maximum allowed perturbation for attribute $j$.

\label{def:cvcbic2}
\end{mydef}

The two highlighted biclusters in Table~\ref{tab:maDataEx2} are the same as the two highlighted biclusters in Table~\ref{tab:maDataEx1}, considering the numerical conversion defined along this subsection. Therefore, this new definition of CVC biclusters completely meets the requirements for mining biclusters in mixed-data matrices.

Remarkably, this new definition of CVC biclusters also meets the monotonicity and anti-monotonicity properties, which is fundamental for enumerative algorithms (as pointed in \cite{VeronezeEtAl2017}). This new definition can also be used to mine biclusters in data matrices formed solely by numerical attributes. For each numerical attribute $j$, $\epsilon_j$ will reflect the range of values assumed by that attribute. Clearly, this new definition can also be used to mine biclusters in data matrices formed solely by categorical attributes.

\section{Class Association Rules}
\label{sec:CAR}

The concepts and metrics for traditional \emph{association mining} \cite{ceglarEtAl2006} are based on binary datasets. So, we will first provide the main concepts and metrics based on the binary case and, after that, we will generalize them to mixed-attribute datasets.

Let $\mathbf{A}_{n \times m}$ be a binary matrix with the row index set $X = \left \{ 1, 2,..., n \right \}$ and the column index set $Y = \left \{ 1, 2, ...,m \right \}$. Each row represents an object, and each column represents an attribute (or \emph{item}, which is the name commonly used in FPM).

\begin{mydef}
A subset $J = \left \{ j_1,..., j_s \right \} \subseteq Y$ is called an \emph{itemset}.
\end{mydef}

Let $I$ be the set of objects that are common to all the items in the itemset $J$, which is given by:
\begin{equation}
I  = \{i \in X | a_{ij} = 1, \; \forall j \in J\}.
\label{eq:support}
\end{equation}

\noindent Notice that the pair $(I,J)$ is a CTV bicluster of 1s.

The \emph{support} of an itemset $J$ is given by
\begin{equation}
sup(J) = |I|,
\label{eq:support}
\end{equation}

\noindent where $|\zeta|$ is the number of elements in the set $\zeta$. The \emph{relative support} of an itemset $J$ is given by

\begin{equation}
rsup(J) = \frac{sup(J)}{n}.
\end{equation}

\begin{mydef}
\noindent An itemset $J$ is a \emph{frequent itemset} if $sup(J) \geq mR$, where $mR$ is a user-defined threshold.
\end{mydef}

\begin{mydef}
\noindent An \emph{association rule} (AR) is an expression in the form $J \Rightarrow H$, where $J$ and $H$ are itemsets and $J \cap H = \emptyset$. $J$ is called the \emph{body} or \emph{antecedent}, and $H$ is called the \emph{head} or \emph{consequent} of the rule.
\end{mydef}

Let us assume that the objects of the data matrix $\mathbf{A}$ are labeled, let $C = \{c_1, c_2, ..., c_k\}$ be the set of possible class labels of the objects, and let $c \in C$.

\begin{mydef}
A \emph{class association rule} (CAR) is an expression of the form $J \Rightarrow c$, where $J$ is an itemset and $c$ is a class label.
\end{mydef}

As we are still only talking about the binary case, a CAR of the type $J \Rightarrow c$ means that the presence of the attributes in $J$ implies class label $c$. For instance, let $\mathbf{A}$ be a matrix whose objects are patients and attributes are symptoms. Let the set of class labels $C$ represents some diseases. Let the itemset $J$ represents the following symptoms $\{fever, nausea, lumbarPain, urethraBurning\}$, and let the class label $c$ represents the disease \emph{Nephritis}. Thus, $J \Rightarrow c$ ($\{fever, nausea, lumbarPain, urethraBurning\} \Rightarrow Nephritis$) means that if a patient has fever, nausea, lumbar pain and urethra burning, then it has Nephritis, with a certain confidence.

Let $c_{I}$ be the set of objects from the data matrix $\mathbf{A}$ with class label $c$, i.e, $c_{I} = \{i \in X | label(i) = c\}$, where $label(.)$ is a function that returns the class label of an object. The \emph{support} of a class label $c$ is given by $sup(c) = |c_{I}|$.

The \emph{support} of a CAR of the type $J \Rightarrow c$ is given by:
\begin{equation}
sup(J \Rightarrow c) = |I \cap c_{I}|.
\end{equation}

\noindent Thus, the \emph{relative support} of a CAR of the type $J \Rightarrow c$ provides an estimate of the probability of the joint occurrence of itemset $J$ and class label $c$:

\begin{equation}
rsup(J \Rightarrow c) = \frac{sup(J \Rightarrow c)}{n} = P(J \wedge c).
\end{equation}

The \emph{confidence} of a CAR of the type $J \Rightarrow c$ is the conditional probability that an object belongs to the class $c$ given that it contains the itemset $J$:

\begin{equation}
conf(J \Rightarrow c) = P(c|J) = \frac{P(J \wedge c)}{P(J)} = \frac{rsup(J \Rightarrow c)}{rsup(J)} = \frac{sup(J \Rightarrow c)}{sup(J)}.
\end{equation}

The completeness of a CAR of the type $J \Rightarrow c$ is given by

\begin{equation}
comp(J \Rightarrow c) = P(J|c) = \frac{P(J \wedge c)}{P(c)} = \frac{rsup(J \Rightarrow c)}{rsup(c)} = \frac{sup(J \Rightarrow c)}{sup(c)}.
\end{equation}

\noindent Thus, completeness is the proportion of instances that are predicted by a CAR of the type $J \Rightarrow c$, while confidence is the fraction of correct predictions made by this CAR.

\emph{Lift} is defined as the ratio of the observed joint probability of $J$ and $c$ to the expected joint probability if they were statistically independent, that is,

\begin{equation}
lift(J \Rightarrow c) = \frac{P(J \wedge c)}{P(J)P(c)} = \frac{rsup(J \Leftarrow c)}{rsup(J)rsup(c)} = \frac{conf(J \Leftarrow c)}{rsup(c)}.
\end{equation}

\noindent One common use of lift is to measure the degree of surprise of a rule. A lift value close to 1 means that the support of a rule is expected considering the support of its components. We usually look for values that are much larger than 1 (i.e., above expectation) or smaller than 1 (i.e., below expectation). Notice that lift is always larger than or equal to the confidence because it is the confidence divided by the consequent's probability.

\emph{Leverage} measures the difference between the observed and expected joint probability of $J$ and $c$ assuming they are independent, that is,

\begin{equation}
leverage(J \Rightarrow c) = P(J \wedge c) - P(J)P(c) = rsup(J \Leftarrow c) - rsup(J)rsup(c).
\end{equation}

\noindent Leverage gives an "absolute" measure of how surprising a rule is. If two rules have the same confidence and lift, the metric leverage indicates which one is stronger.

\subsection{Quantitative Class Association Rules}
\label{ssec:QCAR}

Now, let us think about more general cases where we do not consider only binary attributes. In the literature, these rules are called \emph{quantitative association rules} (QAR) \cite{SrikantAgrawal1996, zhu2009}. In the binary case, we omit the domain of interest of the attributes because they are always equal to 1. For instance, in the itemset $\{fever, nausea, lumbarPain, urethraBurning\}$, we are assuming that our objects of interest are the ones exhibiting these symptoms. So, we could rewrite this itemset as $\{fever\{1\}, nausea\{1\}, lumbarPain\{1\}, urethraBurning\{1\}\}$. In more general cases, each rule must always indicate the range of values of its attributes. For instance, the \emph{quantitative-itemset} $\{Sex\{M\}, Height[1.54,1.62], Smoker\{N\}, SocialClass\{B,C\}\}$ refers to the objects of the dataset that have the attribute \emph{Sex} equal to \emph{M}, the attribute \emph{Height} in the interval $[1.54,1.62]$, the attribute \emph{Smoker} equal to \emph{N}, and the attribute \emph{Social Class} equal to \emph{B} or \emph{C}. Notice that this information is provided by one of the biclusters highlighted in Table~\ref{tab:maDataEx1}: the one composed of the rows \{10, 13, 14, 20\} and attributes \{Sex, Height, Smoker, SocialClass\}.

\begin{mydef}
A \emph{quantitative-itemset} $\mathfrak{J}$ is a set of attributes and their domain of interest, i.e., $\mathfrak{J} = \{j_1 \in D_1, j_2 \in D_2, ..., j_s \in D_s \}$, where $j_\mathfrak{i}$ is an attribute and $D_\mathfrak{i}$ is its domain of interest. If $j_\mathfrak{i}$ is a discrete attribute, $D_\mathfrak{i}$ is a finite set of values; if $j_\mathfrak{i}$ is a continuous attribute, $D_\mathfrak{i}$ is an interval.
\end{mydef}

\begin{mydef}
A \emph{quantitative association rule} (QAR) is an expression of the form $\mathfrak{J} \Rightarrow \mathfrak{H}$, where $\mathfrak{J}$ and $\mathfrak{H}$ are quantitative-itemsets, and the intersection between the attributes of $\mathfrak{J}$ and $\mathfrak{H}$ is empty.
\label{def:qar}
\end{mydef}

\begin{mydef}
A \emph{quantitative class association rule} (QCAR) is an expression of the form $\mathfrak{J} \Rightarrow c$, where $\mathfrak{J}$ is a quantitative-itemset and $c$ is a class label.
\end{mydef}

Thus, a quantitative-itemset is simply a generalization of an itemset, as well as a QAR and a QCAR are generalizations of an AR and a CAR, respectively.

Notice that $\mathbf{A}_{n \times m}$ is now a mixed-data matrix, with each column admitting only numerical or only categorical values. Let $I$ be the set of objects that meets the requirement imposed by the quantitative-itemset $\mathfrak{J}$:

\begin{equation}
I = \{i \in X| a_{ij} \in D, \; \; \forall (j \in D) \in \mathfrak{J} \}.
\end{equation}

The \emph{support} of $\mathfrak{J}$ is given by
\begin{equation}
sup(\mathfrak{J}) = |I|,
\label{eq:support}
\end{equation}
\noindent and it follows that all other metrics previously presented for CARs will be calculated in the same way for QCARs.

Let $J$ be equal to the column indexes of the attributes in $\mathfrak{J}$. Thus, the pair $(I,J)$ is a CVC bicluster. So, a CVC bicluster $(I,J)$ provides all the necessary information to build a quantitative-itemset (that is a component of a QAR, or the antecedent of a QCAR), and vice versa. Notice also that a CVC bicluster is a generalization of a CTV bicluster.

It is important to highlight that quantitative association rules are generally divided into two classes in the literature: frequent rules and distributional rules \cite{zhu2009}. Our definitions are based on \emph{frequent rules} because it is the case that has a direct relation with the biclustering problem. See \cite{zhu2009} for more details about distributional rules.

\subsection{Filters to select significant biclusters from the enumerative solution}

Enumerative algorithms generally return a huge amount of biclusters. In this paper, we are proposing two filters to select a reduced set of significant biclusters from the enumerative solution.

The first filter (\emph{1st-Filter}) is based on FPM metrics to measure the quality of a rule. To mine the biclusters, the user must set the parameter $mR$ in RIn-Close, which is equivalent to the minimum \emph{support} of an itemset (or rule). To discard mined biclusters of low relevance, we rely upon the metrics \emph{confidence} and \emph{lift}. Confidence informs us the fraction of correct predictions, and lift measures the degree of surprise of the rule. Thus, the first step to implement this filter is to build QCARs from the enumerated biclusters. The second and final step is to select the biclusters whose QCARs meet the user-defined thresholds for these two metrics.

A bicluster may contain objects belonging to more than one class in its extent, but we assume that each bicluster represents only one class, the one that holds the majority of the objects in the bicluster's extent. For instance, let the bicluster's extent $I$ be equal to  $I = \{4, 8, 14, 15, 17\}$ and the class labels of these objects be, respectively, $\{0, 0, 1, 0, 1\}$. Then, the class label represented by the bicluster is $0$. For each enumerated bicluster, we build a QCAR based only in the class label that it represents. This rule has the highest confidence among the alternative QCARS that we can build based on this bicluster.

The amount of biclusters selected by the 1st-filter can still be large, so we are proposing a greedy heuristic as a second filter (\emph{2nd-Filter}). Its goal is to select a very small set of representative biclusters, thus allowing manual inspection. The selected biclusters will be exhibited in the form of QCARs, which are highly interpretable.

Algorithm~\ref{alg:greedyHeur} shows the pseudocode of the greedy heuristic. To compute the row-coverage of a bicluster, we consider only the objects belonging to the class represented by that bicluster. For instance, in the previous example, the rows covered by the bicluster considering the represented class are 4, 8, and 15. The row-coverage of a biclustering solution $\mathfrak{B}$ is the union of the row-coverage of its biclusters. The final set of chosen biclusters will have the same row-coverage of the filter input. When more than one bicluster provides the same row-coverage of $\mathfrak{B}'$, the bicluster with the smaller intent is prioritized. For each class, the bicluster with the highest \emph{completeness} will be always selected by this heuristic.

\linespread{1}
\begin{algorithm}
\caption{Greedy Heuristic (2nd-Filter)}
\label{alg:greedyHeur}
\begin{algorithmic}[1]
  \small
  \REQUIRE Biclustering solution $\mathfrak{B}$ 
  \ENSURE Biclustering solution $\mathfrak{B}'$
  \STATE $cov \leftarrow$ row-coverage of $\mathfrak{B}$
  \STATE $aux \leftarrow 0$
  \WHILE{$aux < cov$}
    \STATE Select the bicluster $(I,J)$ from $\mathfrak{B}$ that maximizes the row-coverage of $\mathfrak{B}'$ \COMMENT{In the case of a tie, choose the bicluster with the smallest intent}
    \STATE Insert $(I,J)$ in $\mathfrak{B}'$
    \STATE Remove $(I,J)$ from $\mathfrak{B}$
    \STATE $aux \leftarrow$ row-coverage of $\mathfrak{B}'$
  \ENDWHILE
\end{algorithmic}
\end{algorithm}
\linespread{1.5}

According to Xing \emph{et al.} \cite{XinEtAl2006}, a useful compact pattern set should simultaneously exhibit high significance and low redundancy. These two goals are achieved by our cascade of filters. By means of \emph{1st-Filter}, we select only significant biclusters. And, by means of \emph{2nd-Filter}, which locally maximizes the row-coverage, we select a smaller set of biclusters with low-redundancy, but keeping the same row-coverage of the filter input.

\section{Related Works}
\label{sec:relwork}



In \cite{VandrommeEtAL2016}, the authors presented a biclustering method designed to handle mixed-attribute datasets. This method uses a pre-processing step to simplify the data by means of discretization, and a constructive greedy heuristic to build the biclusters by iteratively adding columns. Their goal was, as expected, to detect CVC biclusters. To the best of Vandromme \emph{et al.}'s knowledge, this was the first method to handle mixed-attribute datasets in the biclustering literature \cite{VandrommeEtAL2016}.

A core principle here is the fact that, when we discretize the numerical attributes, we are no longer looking for biclusters in a mixed-data matrix. Real-valued attributes are each one mapped to a discrete attribute. And even in the case of integer attributes with many distinct values, they are mapped to a smaller set of discrete values. So, after pre-processing, any biclustering algorithm that handles discrete matrices can be used. Discretization may simplify the biclustering task, but it implies loss of information and there is no control on the overall effect of the discretization step on the final results.

In fact, after imposing discretization, we have better proposals than \cite{VandrommeEtAL2016}, especially when we consider the connection between biclustering, FPM and FCA. Notice that we can extract CVC biclusters from quantitative-itemsets (and vice-versa), and (quantitative) association rules are mined from (quantitative-)frequent itemsets. Veroneze \textit{et al.} \cite{VeronezeEtAl2017} also showed that well-known heuristic-based biclustering algorithms can have a poor performance when trying to identify the existing biclusters in a simple and controlled scenario, thus fully favouring the use of efficient enumerative algorithms, such as the ones provided in FPM and FCA literature and the RIn-Close family \cite{VeronezeEtAl2017}.

An approach to mine biclusters from non-binary datasets using traditional FPM and FCA algorithms devoted to binary datasets (such as Apriori \cite{AgrawalSrikant1994}, Charm \cite{ZakiEtAL2002}, or In-Close2 \cite{Andrews2011}) is (1) to discretize the dataset, and (2) to itemize the discrete dataset. Notice that each dataset attribute is an item in the binary case. Basically, the itemization (the second step of the proposed approach) consists in creating a binary dataset from a discrete dataset, without information loss. The first step will necessarily involve some kind of information loss. An item here is a pair $<att,v>$, where $att$ is an attribute (of the original dataset), and $v$ is a discretized value. So, we have as many items as the number of pairs $<att,v>$. Thus, there is a trade-off between faster execution time with fewer discretized values and reduced information loss with more discretized values. Therefore, depending on the nature of the dataset, the user is not totally free to choose the granularity of the discretization.

As far as we known, Srikant \& Agrawal \cite{SrikantAgrawal1996} were the first ones to address the problem of mining quantitative association rules in mixed-attribute datasets. Their proposal discretizes the numerical (quantitative) attributes into partitions or intervals, using equi-depth partitioning. The authors proposed a metric, called partial completeness, to estimate the information loss and help the user to choose the number of intervals. Notice that if a numerical attribute has few distinct values, it does not need to be partitioned in intervals. After the partition in intervals, all the attributes (numerical and categorical) are mapped to positive integers. The next step is the itemization. To solve the problem of not finding rules due to the minimum support, the proposal combines adjacent intervals (or values) of quantitative attributes. Therefore, instead of using a pair $<att,v>$ for an item, the proposals uses a triplet $<att,l, u>$, where $l$ is the lower bound and $u$ is the upper bound. $l = u$ if the attribute is categorical. A single element of the original matrix could be assigned to more than one item (which is similar to the multiple item assignments used in BicPAM \cite{HenriquesMadeira2014}). As a consequence, there is a tendency of an increase in the computational cost and of the occurrence of redundant frequent itemsets (consequently rules). To make the demand for computational resource and the degree of redundancy still worse, the proposal adopts an Apriori-based algorithm to mine the frequent itemsets (Apriori-based algorithms mine all frequent itemsets, not only the closed frequent itemsets).

Let us emphasize that the set of closed frequent itemsets uniquely determines the exact frequency of all frequent itemsets, and it can be orders of magnitude smaller than the set of all frequent itemsets \cite{ZakiEtAL2002}. Moreover, the usage of closed frequent itemsets instead of frequent itemsets drastically reduces the number of rules that have to be presented to the user, without any information loss \cite{LakhalStumme2005}. Charm \cite{ZakiEtAL2002} is an example of an FPM algorithm that mines closed frequent itemsets.

Garcia et al. \cite{GarciaEtAl2010} proposed a multivariate discretization algorithm based on clustering, called \emph{Clustering Based Discretization} (CBD). Only the attributes with higher values of purity (a measure that informs how well the attribute discriminates the classes) are used. So, CBD considers class labels in its discretization routine, being a proposal suitable only for labeled datasets. After the conversion of continuous attributes to discrete ones, the dataset is itemized and a traditional FPM algorithm can be applied.

BicPAM \cite{HenriquesMadeira2014} and BiC2PAM \cite{HenriquesMadeira2016} also relies on discretization, itemization, and the usage of a traditional FPM algorithm to mine the biclusters. BiC2PAM extends BicPAM to incorporate constraints derived from background knowledge in the mining process. BicPAM and BiC2PAM are available in a free bicluster software called BicPAMS \cite{HenriquesEtAl2017}.

BicPAM is a framework that relies on 3 steps: pre-processing (which includes normalization, discretization, itemization, handling of missing values, and tackling varying levels of noise), mining (where some FPM algorithm is used to mine the biclusters), and post-processing (in which the biclusters can be extended, merged and filtered out, among other possibilities). BicPAM makes available three discretization options (each one with key implications on the target solution), and the user can easily incorporate other options into the framework. BicPAM also makes available several FPM algorithms in the mining step, and the user can also incorporate others. To alleviate common drawbacks related to discretization procedures (such as information loss), the user can choose to assign multiple items over a single element, tackling the items-boundary problem. The drawback of this strategy is that it usually generates many redundant biclusters (even when using algorithms to mine closed frequent itemsets), guiding to extra computational cost, and the information loss is still present, though attenuated. For more contributions regarding biclustering based on FPM algorithms, see the survey of Henriques \emph{et al.} \cite{HenriquesEtAl2015}.

The missing values can be simply ignored in methods that rely in itemization to mine the biclusters. Henriques \& Madeira \cite{HenriquesMadeira2014} also proposed the use of additional items, specially handled according to a level of relaxation imposed by the user.

Aiming at bypassing the itemization step, we may resort to enumerative biclustering algorithms that mine CVC biclusters directly from numerical matrices, such as RIn\_Close\_CVC, RIn\_Close\_CVCP and their competitors \cite{VeronezeEtAl2017}. They are able to mine the biclusters from a discretized matrix (that has only integer numbers), thus avoiding itemization. This implies that it is possible to use a more flexible discretization, without restrictions in the arity of an attribute. Additionally, RIn\_Close\_CVCP is a very efficient algorithm, exhibiting a computational cost similar to that of In-Close2.

In conceptual terms, approaches relying on discretization as well as our proposal can control the level of noise inside a bicluster, but not at the same extent. By construction, after defining the discretization policy, approaches relying on discretization are able to guarantee that the level of noise in the mined biclusters will belong to a specific interval. However, they are not able to guarantee, for arbitrary matrices, finding all the biclusters exhibiting a level of noise inside that interval. On the other hand, our proposal guarantees to find all the maximal biclusters exhibiting a level of noise inside a given interval. Therefore, no matter how optimized the discretization policy, any a priori and computationally feasible discretization involves information loss, being not able to compete with online approaches such as ours.

To illustrate this relevant limitation of approaches based on a priori discretization, Table~\ref{tab:rvMatrix} shows an arbitrary example of a real-valued matrix with 10 objects and 3 attributes, and Table~\ref{tab:dMatrix} shows this matrix after discretization using equi-width partitioning with bins of size 0.2. As the matrix of Table~\ref{tab:rvMatrix} was randomly created using a uniform distribution, the equi-width partitioning is a reasonable choice. Notice that the itemization process would produce 15 items, thus this small example indicates the restrictions imposed by the itemization in the arity of the discretization. Using our proposal with minimum number of objects and attributes set to 2, and $\epsilon = 0.2$ for all attributes, we would obtain 12 biclusters, which are listed on Table~\ref{tab:bics_rvMatrix}. Using an FPM or FCA algorithm, such as Charm, in the itemized matrix computed from the matrix of Table~\ref{tab:dMatrix} (with the same restrictions of minimum number of objects and attributes), we would obtain only 7 of these 12 biclusters, being 5 of these 12 biclusters only partially recovered. These 7 biclusters are highlighted in bold in Table~\ref{tab:bics_rvMatrix}. Of course, we could alleviate the information loss using multiple item assignments. However, how many items per element should we assign in a real-world problem? Moreover, multiple item assignment tends to generate redundant biclusters and extra computational cost.

As already mentioned, an adequate discretization process has a significant impact on the quality of the biclusters, and also in the computational cost of the proposals. There are numerous discretization methods available in the literature. Liu \emph{et al.} \cite{LiuEtAl2002} presented a survey of discretization methods and discussed various dimensions in which discretization methods can be categorized. They also gave some guidelines for how to choose a discretization method under various circumstances. However, they stated that the choice of a suitable discretization method is generally a complex matter, largely depending on the demands and particularities of the application. For instance, if the data does not have class information, only unsupervised methods can be applied. They also stated that the availability of parallel computing or computer clusters opens the possibility of using multivariate discretization.

As there is a combinatorial explosion of possibilities for discretization, a fair comparison will require a vast series of experiments and a careful analysis of the context involved, properly addressing pros and cons. Therefore, such an issue is out of the scope of this work and we end the section reminding that no matter the quality of the discretization, information loss will occur. That is why we have conceived our approach to avoid a priori discretization.

\linespread{1}
\begin{table}[!htb]
\begin{minipage}[t]{.47\textwidth}
\caption{Example of a real-valued matrix.}
\label{tab:rvMatrix}
\centering
\footnotesize
\begin{tabular}{crrr}
\toprule
\textbf{\#} & \textbf{1} & \textbf{2} & \textbf{3} \\
\midrule
\textbf{1} & 0.278 & 0.422 & 0.743 \\
\textbf{2} & 0.547 & 0.916 & 0.392 \\
\textbf{3} & 0.958 & 0.792 & 0.655 \\
\textbf{4} & 0.965 & 0.959 & 0.171 \\
\textbf{5} & 0.158 & 0.656 & 0.706 \\
\textbf{6} & 0.971 & 0.036 & 0.032 \\
\textbf{7} & 0.957 & 0.849 & 0.277 \\
\textbf{8} & 0.485 & 0.934 & 0.046 \\
\textbf{9} & 0.800 & 0.679 & 0.097 \\
\textbf{10} & 0.142 & 0.758 & 0.823 \\
\bottomrule
\end{tabular}
\end{minipage}
\begin{minipage}[t]{.53\textwidth}
\caption{Matrix of Table~\ref{tab:rvMatrix} after discretization using equi-width partitioning with bins of size 0.2.}
\label{tab:dMatrix}
\centering
\footnotesize
\begin{tabular}{crrr}
\toprule
\textbf{\#} & \textbf{1} & \textbf{2} & \textbf{3} \\
\midrule
\textbf{1} & 2 & 3 & 4 \\
\textbf{2} & 3 & 5 & 2 \\
\textbf{3} & 5 & 4 & 4 \\
\textbf{4} & 5 & 5 & 1 \\
\textbf{5} & 1 & 4 & 4 \\
\textbf{6} & 5 & 1 & 1 \\
\textbf{7} & 5 & 5 & 2 \\
\textbf{8} & 3 & 5 & 1 \\
\textbf{9} & 4 & 4 & 1 \\
\textbf{10} & 1 & 4 & 5 \\
\bottomrule
\end{tabular}
\end{minipage}
\end{table}
\linespread{1.5}

\linespread{1}
\begin{table}[!htb]
\caption{Biclusters mined from the data matrix of Table~\ref{tab:rvMatrix} using our proposal (to be formally presented in Section~\ref{sec:rinclose}) with minimum number of objects and attributes set to 2, and $\epsilon = 0.2$ for all attributes.}
\label{tab:bics_rvMatrix}
\centering
\footnotesize
\begin{tabular}{ccc}
\toprule
\textbf{\#} & \textbf{Objects} & \textbf{Attributes} \\
\midrule
\textbf{1}   & 1, 5, 10   & 1, 3  \\
\textbf{2}   & \textbf{5, 10}      & \textbf{1, 2}, 3  \\
\textbf{3}   & \textbf{2, 8}       & \textbf{1, 2}  \\
\textbf{4}   & 3, 7, 9    & 1, 2  \\
\textbf{5}   & 7, 9       & 1, 2, 3  \\
\textbf{6}   & 3, \textbf{4, 7}    & \textbf{1, 2}  \\
\textbf{7}   & \textbf{4, 7}       & \textbf{1, 2}, 3  \\
\textbf{8}   & \textbf{4, 6}, 9    & \textbf{1, 3}  \\
\textbf{9}   & 4, 7, 9    & 1, 3  \\
\textbf{10}  & \textbf{3, 5}, 10   & \textbf{2, 3}  \\
\textbf{11}  & \textbf{2, 7}       & \textbf{2, 3}  \\
\textbf{12}  & \textbf{4, 8}       & \textbf{2, 3}  \\
\bottomrule
\end{tabular}
\end{table}
\linespread{1.5}

\section{The Extended Version of RIn-Close\_CVC}
\label{sec:rinclose}

Veroneze \emph{et al.} \cite{VeronezeEtAl2017} proposed an algorithm to enumerate all maximal CVC biclusters in numerical datasets, named RIn-Close\_CVC. From now on, we will call it RIn-Close for simplicity. RIn-Close \cite{VeronezeEtAl2017} is based on Definition~\ref{def:cvcbic} for CVC biclusters. In this section, we will generalize RIn-Close \cite{VeronezeEtAl2017} to prepare this biclustering algorithm to enumerate CVC biclusters in mixed-data matrices. Thus, this new version will be based on Definition~\ref{def:cvcbic2}, which is a generalization of Definition~\ref{def:cvcbic}. Strictly numerical datasets and strictly categorical datasets can also be treated by this extended version of RIn-Close.

The previous version of RIn-Close \cite{VeronezeEtAl2017} was not prepared to enumerate biclusters directly from a dataset with missing values. Some strategies to handle missing values were proposed for the RIn-Close family of algorithms \cite{VeronezeEtAl2017}. The simplest one is to remove the rows and/or columns (usually the ones with smaller dimension) containing missing values, at the cost of information loss. Another simple strategy is the previous estimation of the missing values using some imputation technique of the literature. The problem with this approach is that it generally will introduce additional noise to the dataset, which may significantly reduce the biclusters' homogeneity, thus promoting unnecessary bicluster partitioning. Essentially, a single large original bicluster with some missing elements may be recovered as dozens of smaller biclusters, possibly with a high overlap among them \cite{OliveiraEtAl2015}.

Here, RIn-Close will be extended to mine biclusters directly from datasets with missing values. We will look for biclusters in the regions of the dataset without missing values, ignoring the regions with missing values. Thus, the sparser the matrix, the smaller the portion of the matrix to be mined. Our approach to deal with missing data has a low computational cost, avoids information loss, and does not introduce additional noise into the dataset \cite{Veroneze2016}, being consistent with highly competitive approaches in the literature \cite{HenriquesMadeira2014}.

Besides the incorporation of these new features, this new version of RIn-Close keeps the four key properties of the original proposal \cite{VeronezeEtAl2017}: efficiency, completeness, correctness, and non-redundancy. Also, it has the same worst-case time complexity.

Algorithms~\ref{alg:rinclose} to \ref{alg:ComputeRM} present the pseucode of this new version of RIn-Close and of its main functions. The proposed new features are highlighted in red. Note that with few and simple modifications, we are able to reach our goal: provide an algorithm to enumerate all maximal biclusters in mixed-data (or strictly numerical / categorical) matrices with (or without) missing values.

Firstly, we create the \emph{supremum} bicluster $(I,J)$ in Algorithm~\ref{alg:rinclose}, which contains all rows of the dataset in its \emph{extent} (set of rows of the bicluster), and no column in its \emph{intent} (set of columns of the bicluster). From the supremum, all other biclusters will be mined recursively. This strategy was already adopted by In-Close2 \cite{Andrews2011}, which is the enumerative algorithm for binary datasets that, after generalizations and extensions, gave rise to the RIn-Close family of algorithms \cite{VeronezeEtAl2017}.

\linespread{1}
\begin{algorithm}
\caption{RIn-Close}
\label{alg:rinclose}
\begin{algorithmic}[1]
  \small
	\REQUIRE Data matrix $\mathbf{A}_{n \times m}$, minimum number of rows $mR$, minimum number of columns $mC$, vector with the user-defined maximum perturbation for each attribute $\mathbf{\epsilon}$
	\ENSURE Biclustering solution $\mathfrak{B}$
	\STATE $y \leftarrow 1$ \COMMENT{index of the initial attribute}
	\STATE $I \leftarrow \{ 1, 2,..., n \}$ \COMMENT{extent - set of rows of the bicluster}
	\STATE $J \leftarrow \{ \}$ \COMMENT{intent - set of columns of the bicluster}
	\STATE $\Gamma \leftarrow \{ \}$ \COMMENT{set of rows to check the row-maximality of the descendants}
	\STATE ComputeBiclustersFrom($(I,J),y, \Gamma$)
\end{algorithmic}
\end{algorithm}
\linespread{1.5}

\linespread{1}
\begin{algorithm}
\caption{ComputeBiclustersFrom}
\label{alg:ComputeBiclustersFrom}
\begin{algorithmic}[1]
  \small
  \REQUIRE Bicluster $(I,J)$ to be closed, current attribute $y$, set of rows to check the row-maximality of the descendants $\Gamma$
  \FOR{$j \leftarrow y$ to $m$}
	  \IF{$j \notin J$}
		  \IF{$\max_{i \in I}(a_{ij}) - \min_{i \in I}(a_{ij}) \leq {\color{red}\epsilon_j}$ {\color{red}\AND $a_{ij} \neq mv, \forall i \in I$}}
			  \STATE $J \leftarrow J \cup \{j\}$
			\ELSE
			  \STATE Compute the possible new extents \COMMENT{Eq.~\ref{eq:rinc_cvc_compExt}}
			  \FOR{each possible new extent $G$}
					\IF{$|G| \geq mR$ \AND $G \notin ST$ \AND $G$ is canonical \AND $G$ is row-maximal}
					    \STATE Sort the row indexes in $G$
					    \STATE Insert $G$ in the symbol table $ST$
						\STATE $\Omega \leftarrow ComputeRM(G, j, \Gamma)$
						\STATE PutInQueue($G, j, \Omega$)
					\ENDIF					
				\ENDFOR
			\ENDIF
		\ENDIF
	\ENDFOR
	\IF{$|J| \geq mC$}
	  \STATE Store the bicluster $(I,J)$ in the solution $\mathfrak{B}$
	\ENDIF
	\WHILE{GetFromQueue($G, j, \Omega$)}
	  \STATE $H \leftarrow J \cup \{j\}$
		\STATE ComputeBiclustersFrom($(G,H),j+1, \Omega$)
	\ENDWHILE
\end{algorithmic}
\end{algorithm}
\linespread{1.5}

In Algorithm~\ref{alg:ComputeBiclustersFrom}, each bicluster $(I,J)$ is closed, i.e., its intent is completed with all possible columns for the extent $I$ (line 4). The expression $a_{ij} \neq mv$ means: the element $a_{ij}$ is not a missing value. If the attribute $j$ is not an inherited attribute and it cannot be added to the intent $J$, the possible new extents are computed (line 6). Given that $I$ is the current extent, and $j$ is the current attribute, the possible new extents are given by
\begin{equation}
	\{G | [G \subseteq I] \; \wedge \; [\max_{i \in G}(\{a_{ij}\}) - \min_{i \in G}(\{a_{ij}\}) \leq {\color{red}\epsilon_j}] {\color{red} \; \wedge \; [a_{ij} \neq mv, \forall i \in G] } \; \wedge \; [G \; \mathrm{is \; maximal}]\}.
	\label{eq:rinc_cvc_compExt}
\end{equation}

\noindent It is easily achieved by ordering the values of the data matrix $\mathbf{A}$ in rows $I$ and column $j$. The user should use a large number to represent the missing values ($mv$), so the missing values will be at the bottom of the list, and this whole portion of the list can be ignored. If a new possible extent $G$ passes the verifications of line 8, then it will give rise to a new bicluster with extent $G$.

Letting $J$ be the current intent, and $j$ be the current attribute, a possible new extent $G$ of a bicluster is not canonical if

\begin{equation}
    \exists k \in Y \setminus J | [k < j] \: \wedge \: [\max_{i \in G}(a_{ik}) - \min_{i \in G}(a_{ik}) \leq {\color{red}\epsilon_k}] {\color{red}\: \wedge \: [a_{ik} \neq mv, \forall i \in G]},
	\label{eq:rinc_cvc_iscan}
\end{equation} 

\noindent i.e., if there is an attribute $k < j$ that we can add to the bicluster $(G, J)$ and it remains a valid CVC bicluster.

Letting $J$ be the current intent, $j$ be the current attribute, $H = J \cup \{j\}$, and $\Gamma$ be the set of rows that must be checked to verify the row-maximality, a possible new extent $G$ is not row-maximal if there is an object $g \in \Gamma$ that we can add to the bicluster $(G,H)$ and it remains a valid CVC bicluster, i.e.,

\begin{equation}
  \exists g \in \Gamma | [\max_{i \in \{G \cup \{g\}\}}(a_{ik}) - \min_{i \in \{G \cup \{g\}\}}(a_{ik}) \leq {\color{red}\epsilon_k}] {\color{red} \: \wedge \: [a_{gk} \neq mv]}, \forall \; k \in H. 
	\label{eq:cvc_ismaximal}
\end{equation}

Besides the canonicity and the row-maximality verifications, we also verify if the possible new extent $G$ does not belong to a symbol table $ST$. This verification is based on the fact that two distinct CVC biclusters must have two distinct extents in order to be maximal. So, to avoid redundant maximal biclusters, the extents that have already been generated are tracked using an efficient symbol table implementation, such as hash tables (HTs) or balanced search trees (BSTs).

From these verifications, the only one that was inspired by In-Close2 is the canonicity test, which was originally proposed in \cite{Kuznetsov1996}. Clearly, the canonicity test was generalized by Veroneze \emph{et al.} \cite{VeronezeEtAl2017} to deal with numerical datasets. The other two verifications were already part of the previous version of RIn-Close. Here, we are just updating these three verifications to accomplish the new features of RIn-Close.

\linespread{1}
\begin{algorithm}
\caption{ComputeRM}
\label{alg:ComputeRM}
\begin{algorithmic}[1]
  \small
  \REQUIRE new extent $G$, current attribute $j$, set of rows to check the row-maximality $\Gamma$
  \ENSURE new set of rows to check the row-maximality $\Omega$
  \STATE $\mathbf{v} \leftarrow \{a_{ij}\}_{i \in G}$
  \STATE $\mathbf{v} \leftarrow sort(\mathbf{v})$ \COMMENT{ascending order}
  \STATE $p1 \leftarrow \mathbf{v}_{mR}$ \COMMENT{pivot value 1}
  \STATE $p2 \leftarrow \mathbf{v}_{|\mathbf{v}| - mR + 1}$ \COMMENT{pivot value 2}
  \STATE $\Omega \leftarrow \Gamma \cup \{i \in X \setminus G| \; [[p1 - a_{ij} \leq {\color{red}\epsilon_j}] \;  \vee \; [a _{ij} -p2 \leq {\color{red}\epsilon_j}]] {\color{red}\; \wedge \; [a_{ij} \neq mv]} \}$
\end{algorithmic}
\end{algorithm}
\linespread{1.5}

\begin{figure}
  \centering 
  \includegraphics[trim=2cm 14.5cm 13cm 2.5cm, clip, scale=0.65]{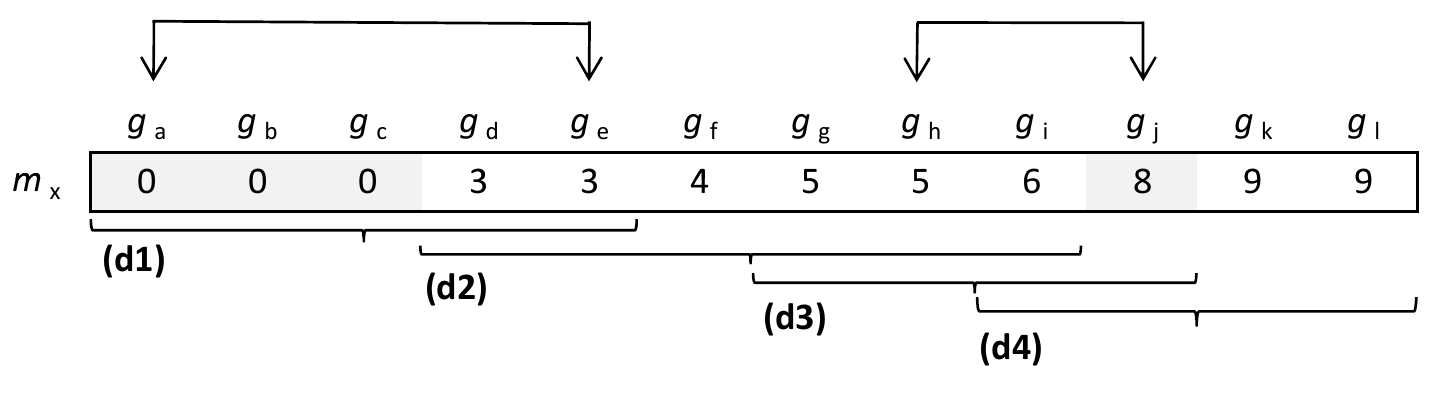}
  \caption{Example of how to find $RM$ (considering $\epsilon = 3$ and $mR = 2$). Extracted from \cite{VeronezeEtAl2017}.}
  \label{fig:RM}
\end{figure}

To explain the function $ComputeRM$ of Algorithm~\ref{alg:ComputeRM}, let us see the example in Figure~\ref{fig:RM}, which considers $\epsilon = 3$ and $mR = 2$. Suppose that $m_x$ is the current attribute and $I = \{g_a, g_b, ..., g_l\}$ is the current extent. So, we have four new possible extents: (d1), (d2), (d3) and (d4). To exemplify, let us compute the set $\Omega$ for (d2). Let us suppose that $\Gamma= \{\}$. The pivot elements are $g_e$ and $g_h$ because they are the $mR$-$th$ first and last elements of (d2), respectively. Their values are $g_e = 3$ and $g_h = 5$. Rows with values greater than or equal to 0 ($g_e - \epsilon$) or less than or equal to 8 ($g_h + \epsilon$) must comprise $\Omega$, so $\Omega = \Gamma \cup \{g_a, g_b, g_c, g_j\} = \{g_a, g_b, g_c, g_j\}$.

This new version of RIn-Close has the same worst-case time complexity of its previous version: $O(kmn(mn + x))$, where $k$ is the number of enumerated biclusters, and $x$ is the worst-case time of searching in the symbol table, so $x = O(\log k)$ for BSTs and $x = O(k)$ for HTs. But HTs have a much better computational cost on average: $O(1)$. For this reason, our RIn-Close implementation uses a HT.

One detail we would like to comment is that we can abort the closure of a bicluster if, even adding all remaining attributes to its intent, it will not meet the minimum number of columns $mC$ (therefore, its next
descendants will not meet the minimum number of columns $mC$ as well). Although this restriction can be checked only during the closure of a bicluster, it will also prune the search space and save computational resources because ($i$) it stops the construction of a bicluster that will be discarded later, given that it does not meet the restriction $mC$, and ($ii$) it avoids generating descendants that will not meet the
restriction $mC$ as well \cite{VeronezeEtAl2017}. This aspect was omitted from Algorithm~\ref{alg:ComputeBiclustersFrom} for the purpose of emphasizing the main steps.

\section{Experimental Results}
\label{sec:exp}

This section describes the datasets used in our experiments and presents the results, followed by an extensive discussion of the main achievements.

\subsection{Description of the datasets}

Table~\ref{tab:datasets} briefly describes the datasets used in our experiments, extracted from the UCI Repository \cite{Lichman2013}. Their attributes are outlined in Tables~\ref{tab:dat_acute} to \ref{tab:dat_zoo}, together with the maximum perturbation $\epsilon$ for each attribute, to be used by the RIn-Close algorithm when mining the biclusters. The attributes are labeled as R (real-valued), I (integer), O (ordinal), or N (nominal). Table~\ref{tab:dataLabels} contains the description of the class labels associated with all the datasets.

These datasets were chosen because they came with the description of all the attributes and class labels. So, biclustering results can be easily interpreted. Given that the number of attributes is not so high, we can easily illustrate how the biclusters guide to interpretative models. Similarly, we can provide QCARs that are able to properly discriminate the class labels.

For the nominal attributes, $\epsilon = 0$ is the only choice that makes sense. For the other types of attributes, we set this parameter based on trial-and-error. Our goal was to provide a biclustering solution with a good coverage of the dataset instances. The choice of the parameter $\epsilon$ for real-valued attributes and for integer attributes with many distinct values were assisted by the bin sizes, returned by the function \emph{histcounts} of MATLAB R2015a.

There are datasets with balanced and unbalanced classes, with 2 to 7 distinct class labels. Most of the datasets are unbalanced. There are class labels with very few instances, for instance, the class label 5 of the Zoo dataset has only 4 instances. The dataset Acute has two decision variables, and we will provide results for both of them. Thus, we are exploring several scenarios to show to the user the biclustering capability to provide interesting rules.

\linespread{1}
\begin{table}[]
\footnotesize
\centering
\caption{Numerical aspects of the datasets.}
\label{tab:datasets}
\begin{tabular}{rlrrrll}
\toprule
\textbf{\#} & \textbf{Name} & \textbf{\# rows} & \textbf{\# columns} & \textbf{\# labels} & \textbf{$mv$} & \textbf{Description}\\
\midrule
1 & Acute \cite{CzerniakEtAl2003} & 120  & 6  & 2 & no  & Acute Inflammations \\
2 & Car    & 1728 & 6  & 4 & no  & Car Evaluation \\
3 & Heart  & 270  & 13 & 2 & no  & Heart Disease \\
4 & Voting & 435  & 16 & 2 & yes & Congressional Voting Records \\
5 & Zoo    & 101  & 16 & 7 & no  & Zoo Animals \\
\bottomrule
\end{tabular}
\end{table}
\linespread{1.5}

\linespread{1}
\begin{table}[]
\footnotesize
\centering
\caption{Description of the attributes in the Acute dataset.}
\label{tab:dat_acute}
\begin{tabular}{rllclr}
\toprule
\textbf{\#} & \textbf{Name} & \textbf{Description} & \textbf{Type} & \textbf{Domain} & \textbf{$\epsilon$}\\
\midrule
1           & temperature     & Temperature of patient                               & R             & $[35.5, 41.5]$ & 2.4              \\
2           & nausea          & Occurrence of nausea                                 & N             & \{yes, no\}      & 0.0                \\
3           & lumbarPain      & Lumbar pain                                          & N             & \{yes, no\}      & 0.0                \\
4           & urinePushing    & Urine pushing (continuous need for urination)        & N             & \{yes, no\}      & 0.0                \\
5           & micturitionPain & Micturition pains                                    & N             & \{yes, no\}      & 0.0                \\
6           & urethraBurning  & Burning of urethra, itch, swelling of urethra outlet & N             & \{yes, no\}      & 0.0 \\
\bottomrule
\end{tabular}
\end{table}
\linespread{1.5}

\linespread{1}
\begin{table}[]
\footnotesize
\centering
\caption{ Description of the attributes in the Car dataset.}
\label{tab:dat_car}
\begin{tabular}{rllclr}
\toprule
\textbf{\#} & \textbf{Name} & \textbf{Description} & \textbf{Type} & \textbf{Domain} & \textbf{$\epsilon$}\\
\midrule
1 & buying  & Buying price                          & O & \{v-high, high, med, low\} & 0 \\
2 & maint   & Maintenance price                     & O & \{v-high, high, med, low\} & 1 \\
3 & doors   & Number of doors                       & O & \{2, 3, 4, 5-more\}        & 1 \\
4 & persons & Capacity in terms of persons to carry & O & \{2, 4, more\}             & 0 \\
5 & lugBoot & Size of the luggage boot              & O & \{small, med, big\}        & 0 \\
6 & safety  & Estimated safety of the car           & O & \{low, med, high\}         & 0 \\
\bottomrule
\end{tabular}
\end{table}
\linespread{1.5}

\linespread{1}
\begin{table}[]
\footnotesize
\centering
\caption{ Description of the attributes in the Heart dataset.}
\label{tab:dat_heart}
\begin{tabular}{rlp{5cm}cp{5cm}r}
\toprule
\textbf{\#} & \textbf{Name} & \textbf{Description} & \textbf{Type} & \textbf{Domain} & \textbf{$\epsilon$}\\
\midrule
1  & age          & Age                                                & I & \{29, 30, ..., 77\}                                                                                                      & 4.0   \\
2  & sex          & Sex                                                & N & \{female, male\}                                                                                                       & 0.0   \\
3  & chestPain    & Chest pain type                                    & N & \{typical angina, atypical angina, non-anginal pain, asymptomatic\}                                                    & 0.0   \\
4  & bloodPres    & Resting blood pressure                             & R & {[}94, 200{]}                                                                                                          & 10.0  \\
5  & chol         & serum cholestoral in mg/dl                         & R & {[}126, 564{]}                                                                                                         & 30.0  \\
6  & fastBSugar   & fasting blood sugar \textgreater 120 mg/dl         & N & \{yes, no\}                                                                                                            & 0.0   \\
7  & electro      & resting electrocardiographic results               & N & \{normal, having ST-T wave abnormality, showing probable or definite left ventricular hypertrophy by Estes' criteria\} & 0.0   \\
8  & heartRate    & maximum heart rate achieved                        & R & {[}71, 202{]}                                                                                                         & 10.0  \\
9  & exercIAngina & exercise induced angina                            & N & \{yes, no\}                                                                                                            & 0.0   \\
10 & oldpeak      & ST depression induced by exercise relative to rest & R & {[}0, 6.2{]}                                                                                                           & 0.5 \\
11 & slope        & the slope of the peak exercise ST segment          & O & \{upsloping, flat, downsloping\}                                                                                       & 0.0   \\
12 & vesselsColor & number of major vessels colored by flourosopy      & I & \{0, 1, 2, 3\}                                                                                                         & 0.0   \\
13 & thal         & thal                                               & N & \{normal, fixed defect, reversable defect\}  & 0.0 \\
\bottomrule
\end{tabular}
\end{table}
\linespread{1.5}

\linespread{1}
\begin{table}[]
\footnotesize
\centering
\caption{ Description of the attributes in the Voting dataset.}
\label{tab:dat_voting}
\begin{tabular}{rllclr}
\toprule
\textbf{\#} & \textbf{Name} & \textbf{Description} & \textbf{Type} & \textbf{Domain} & \textbf{$\epsilon$}\\
\midrule
1  & hInfants        & handicapped-infants                    & N    & \{yes, no\} & 0       \\
2  & wProject        & water-project-cost-sharing             & N    & \{yes, no\} & 0       \\
3  & budgetRes       & adoption-of-the-budget-resolution      & N    & \{yes, no\} & 0       \\
4  & physicianFF     & physician-fee-freeze                   & N    & \{yes, no\} & 0       \\
5  & ES-aid          & el-salvador-aid                        & N    & \{yes, no\} & 0       \\
6  & rgSchools       & religious-groups-in-schools            & N    & \{yes, no\} & 0       \\
7  & antiSatelliteTT & anti-satellite-test-ban                & N    & \{yes, no\} & 0       \\
8  & aidNicaraguaC   & aid-to-nicaraguan-contras              & N    & \{yes, no\} & 0       \\
9  & mxMissile       & mx-missile                             & N    & \{yes, no\} & 0       \\
10 & immigration     & immigration                            & N    & \{yes, no\} & 0       \\
11 & sfCorpCut       & synfuels-corporation-cutback           & N    & \{yes, no\} & 0       \\
12 & eduSpending     & education-spending                     & N    & \{yes, no\} & 0       \\
13 & superfundRS     & superfund-right-to-sue                 & N    & \{yes, no\} & 0       \\
14 & crime           & crime                                  & N    & \{yes, no\} & 0       \\
15 & dutyFree        & duty-free-exports                      & N    & \{yes, no\} & 0       \\
16 & admSA           & export-administration-act-south-africa & N    & \{yes, no\} & 0 \\
\bottomrule
\end{tabular}
\end{table}
\linespread{1.5}

\linespread{1}
\begin{table}[]
\footnotesize
\centering
\caption{ Description of the attributes in the Zoo dataset.}
\label{tab:dat_zoo}
\begin{tabular}{rlclr}
\toprule
\textbf{\#} & \textbf{Name}  & \textbf{Type} & \textbf{Domain} & \textbf{$\epsilon$}\\
\midrule
1  & hair     & N    & \{yes, no\}     & 0       \\
2  & feathers & N    & \{yes, no\}     & 0       \\
3  & eggs     & N    & \{yes, no\}     & 0       \\
4  & milk     & N    & \{yes, no\}     & 0       \\
5  & airborne & N    & \{yes, no\}     & 0       \\
6  & aquatic  & N    & \{yes, no\}     & 0       \\
7  & predator & N    & \{yes, no\}     & 0       \\
8  & toothed  & N    & \{yes, no\}     & 0       \\
9  & backbone & N    & \{yes, no\}     & 0       \\
10 & breathes & N    & \{yes, no\}     & 0       \\
11 & venomous & N    & \{yes, no\}     & 0       \\
12 & fins     & N    & \{yes, no\}     & 0       \\
13 & legs     & I    & \{0,2,4,5,6,8\} & 0       \\
14 & tail     & N    & \{yes, no\}     & 0       \\
15 & domestic & N    & \{yes, no\}     & 0       \\
16 & catsize  & N    & \{yes, no\}     & 0  \\
\bottomrule
\end{tabular}
\end{table}
\linespread{1.5}

\linespread{1}
\begin{table}[]
\footnotesize
\centering
\caption{Class Labels of each dataset.}
\label{tab:dataLabels}
\begin{tabular}{cp{12cm}r}
\toprule

\multicolumn{3}{c}{\textbf{Acute - 1st decision variable}} \\
\midrule
\textbf{Label} & \textbf{Description} & \textbf{\# instances} \\
\midrule
0 & No inflammation of urinary bladder & 61 \\
1 & Inflammation of urinary bladder    & 59 \\
\midrule

\multicolumn{3}{c}{\textbf{Acute - 2nd decision variable}} \\
\midrule
\textbf{Label} & \textbf{Description} & \textbf{\# instances} \\
\midrule
0 & No nephritis of renal pelvis origin  & 70 \\
1 & Nephritis of renal pelvis origin     & 50 \\
\midrule

\multicolumn{3}{c}{\textbf{Car}} \\
\midrule
\textbf{Label} & \textbf{Description} & \textbf{\# instances} \\
\midrule
1 & Unacceptable & 1210 \\
2 & Acceptable   & 384 \\
3 & Good         & 69 \\
4 & Very Good    & 65 \\
\midrule

\multicolumn{3}{c}{\textbf{Heart}} \\
\midrule
\textbf{Label} & \textbf{Description} & \textbf{\# instances} \\
\midrule
0 & Absence of heart disease   & 150 \\
1 & Presence of heart disease  & 120 \\
\midrule

\multicolumn{3}{c}{\textbf{Voting}} \\
\midrule
\textbf{Label} & \textbf{Description} & \textbf{\# instances} \\
\midrule
0 & Republicans  & 168 \\
1 & Democrats    & 267 \\
\midrule

\multicolumn{3}{c}{\textbf{Zoo}} \\
\midrule
\textbf{Label} & \textbf{Description} & \textbf{\# instances} \\
\midrule
1 & aardvark, antelope, bear, boar, buffalo, calf, cavy, cheetah, deer, dolphin, elephant, fruitbat, giraffe, girl, goat, gorilla, hamster,hare, leopard, lion, lynx, mink, mole, mongoose, opossum, oryx, platypus, polecat, pony, porpoise, puma, pussycat, raccoon, reindeer, seal, sealion, squirrel, vampire, vole, wallaby, wolf & 41 \\
2 & chicken, crow, dove, duck, flamingo, gull, hawk, kiwi, lark, ostrich, parakeet, penguin, pheasant, rhea, skimmer, skua, sparrow, swan, vulture, wren & 20 \\
3 & pitviper, seasnake, slowworm, tortoise, tuatara & 5  \\
4 & bass, carp, catfish, chub, dogfish, haddock, herring, pike, piranha, seahorse, sole, stingray, tuna & 13 \\
5 & frog, frog, newt, toad & 4  \\
6 & flea, gnat, honeybee, housefly, ladybird, moth, termite, wasp & 8  \\
7 & clam, crab, crayfish, lobster, octopus, scorpion, seawasp, slug, starfish, worm  & 10 \\
\bottomrule

\end{tabular}
\end{table}
\linespread{1.5}

\subsection{Parameter setting}

To enumerate the biclusters, we set the minimum number of rows and the minimum number of columns to $mR = 5$ and $mC = 1$, respectively, for all datasets but Zoo, for which we use $mR = 3$ due to the existence of a class with only 4 instances. The $\epsilon$ value associated with each attribute is presented in Tables~\ref{tab:dat_acute} to \ref{tab:dat_zoo}.

The parameters of the 1st-Filter are established as follows. The minimum confidence is set to 0.95, and the minimum distance from 1 for the lift metric was set to 0.2. The only exception is again the Zoo dataset, where the minimum confidence is set to 1.00 due to the easiness with which rules with high row-coverage are found. The second filter does not have user-defined parameters.

\subsection{Results and Discussion}

Table~\ref{tab:results} summarizes the results of the biclustering solutions returned by RIn-Close, before and after the application of the cascade of two filters. As already explained, \emph{\% row-coverage} considers only the objects from the class label represented by each bicluster.

This is the only part of the experiments that would admit a comparison with contenders in the literature. However, given that all the existing contenders are based on discretization of the numerical attributes, the discussion raised at the end of Section~\ref{sec:relwork}, supported by an illustrative example, provides sufficient evidence that any discretization-based approach would not achieve a biclustering solution capable of overcoming the results of Table~\ref{tab:results}, given that they are not able to guarantee enumerating all maximal biclusters.

As we have already mentioned, an enumerative algorithm may return a huge amount of biclusters. For instance, RIn-Close mined $189,785$ biclusters in the Voting dataset. Our 1st-Filter, which is based on FPM metrics, was able to obtain a reduction in the amount of biclusters from $40\%$ up to $90\%$. So, it was very effective in selecting a reduced subset of significant biclusters. Nonetheless, the number of biclusters is still far above what could be manually inspected. Now, making use of the 2nd-Filter (the greedy heuristic), we were able to select a minimum of 4 to a maximum of 54 biclusters, depending on the dataset.

By construction, the row-coverage of both filters is the same. The 1st-Filter did not have a great impact on the row-coverage of the biclustering solution. So, even keeping only the biclusters with high confidence and lift, the biclustering solutions are comprising most of the objects of each dataset. The largest reduction occurred for the dataset Car, being of $13\%$. For the other datasets, the reduction was absent or insignificant, even with our choices being quite strict for the user-define thresholds of the 1st-Filter. The smaller the parameter $mR$ (of RIn-Close) and the more relaxed the user-defined thresholds of the 1st-Filter, the greater the coverage. On the other hand, relaxed user-defined thresholds also implies a stronger need for the opinion of an expert to determine the relevance of a bicluster.

As we are dealing with datasets with few attributes, the filters did not have a great impact on the column-coverage. An exception was the results for the second decision variable of Acute dataset: we need only 3 attributes to determine the presence or absence of \emph{Inflammation of urinary bladder}. In case studies characterized by the existence of much more attributes, not considered here, the results tend to be different. For instance, in \cite{Veroneze2016}, RIn-Close was used to analyse a numerical dataset with $2,308$ attributes (genes). A filter similar to the 1st-Filter selected $641$ genes, which means a column-coverage reduction of more than $70\%$. And a filter similar to the 2nd-Filter selected only $62$ of these $641$ genes. This is a promising practical tendency, given that, under the presence of a high number of attributes, being able to automatically select a small subset of relevant attributes is high desirable in biosciences and other related areas.

\linespread{1}
\begin{table}[]
\footnotesize
\centering
\caption{Biclustering Results.}
\label{tab:results}
\begin{tabular}{lrrr}
\toprule

\multicolumn{4}{c}{\textbf{Acute - 1st decision variable}} \\
\midrule
    & \textbf{Original} & \textbf{1st Filter} & \textbf{2nd-Filter} \\
\midrule
\textbf{\# of biclusters}    & 172      & 54       & 4        \\
\textbf{\% row-coverage}     & 100.00   & 100.00   & 100.00   \\
\textbf{\% column-coverage}  & 100.00   & 100.00   &  83.33    \\
\midrule

\multicolumn{4}{c}{\textbf{Acute - 2nd decision variable}} \\
\midrule
    & \textbf{Original} & \textbf{1st-Filter} & \textbf{2nd-Filter} \\
\midrule
\textbf{\# of biclusters}    & 172      & 66       & 4        \\
\textbf{\% row-coverage}     & 100.00   & 100.00   & 100.00   \\
\textbf{\% column-coverage}  & 100.00   & 100.00   &  50.00    \\
\midrule

\multicolumn{4}{c}{\textbf{Car}} \\
\midrule
    & \textbf{Original} & \textbf{1st-Filter} & \textbf{2nd-Filter} \\
\midrule
\textbf{\# of biclusters}    & 4,147     & 1,940     & 54       \\
\textbf{\% row-coverage}     &  98.67   &  85.01   &  85.01   \\
\textbf{\% column-coverage}  & 100.00   & 100.00   & 100.00    \\
\midrule

\multicolumn{4}{c}{\textbf{Heart}} \\
\midrule
    & \textbf{Original} & \textbf{1st-Filter} & \textbf{2nd-Filter} \\
\midrule
\textbf{\# of biclusters}    & 82,150    & 15,808    & 38       \\
\textbf{\% row-coverage}     & 100.00   &  99.63   &  99.63   \\
\textbf{\% column-coverage}  & 100.00   & 100.00   & 100.00    \\
\midrule

\multicolumn{4}{c}{\textbf{Voting}} \\
\midrule
    & \textbf{Original} & \textbf{1st-Filter} & \textbf{2nd-Filter} \\
\midrule
\textbf{\# of biclusters}    & 189,785   & 109,873    & 13       \\
\textbf{\% row-coverage}     &  99.77   &  99.08   &  99.08   \\
\textbf{\% column-coverage}  & 100.00   & 100.00   &  87.50    \\
\midrule

\multicolumn{4}{c}{\textbf{Zoo}} \\
\midrule
    & \textbf{Original} & \textbf{1st-Filter} & \textbf{2nd-Filter} \\
\midrule
\textbf{\# of biclusters}    & 4,429    & 346     & 9       \\
\textbf{\% row-coverage}     & 100.00  & 100.00  & 100.00   \\
\textbf{\% column-coverage}  & 100.00  & 100.00  & 100.00    \\
\bottomrule

\end{tabular}
\end{table}
\linespread{1.5}

Tables~\ref{tab:rulesAcute} to \ref{tab:rulesZoo} show, for each dataset, the QCARs directly extracted from the biclusters select by the cascade of two filters, after enumeration of all existing biclusters. Notice the interpretative power of the QCARs when compared to the corresponding crude biclusters.

Table~\ref{tab:rulesAcute} shows the rules for the dataset Acute. For its first decision variable, we can say that the main rule is the third one, given that it has a very high completeness and leverage. Almost all the patients with inflammation of urinary bladder presented urine pushing (continuous need for urination) and micturition pains. The variable indicating continuous need for urination appears in other two rules (the second and fourth rules). It is a strong indicative that this variable is important to determine the presence or absence of inflammation of urinary bladder. The difference between the second and fourth rules are the presence or absence of urine pushing, both of them indicating the absence of burning of urethra, itch, or swelling of urethra outlet. This set of rules show us that all patients with inflammation of urinary bladder had urine pushing, but there are some patients without inflammation of urinary bladder that also presented urine pushing. In the first rule, the attribute that indicates the presence or absence of micturition pains appears again. In this case, the absence of micturition pains and nauseas, together with the presence of lumbar pain, are indicating the absence of inflammation of urinary bladder.

Now, let us analyse the rules of the second decision variable of the Acute dataset. We note that the third and fourth rules, that indicate presence of nephritis of renal pelvis origin, overlap and could be rewritten as a single rule: temperature[37.90,41.50], lumbarPain\{yes\} $\Rightarrow$ 1. It happens because of the user-defined parameter $\epsilon$ for the attribute temperature (see Table~\ref{tab:dat_acute}). To avoid this event, we figure out two possibilities. The first one is to run the enumerative algorithm few times with different settings for the vector $\epsilon$, producing a set of biclustering solutions. So, we can select relevant biclusters from this pool of solutions. The second possibility is to post-process the rules to group the ones that overlaps or are adjacent. This second option is commonly used by the algorithms that discretize the dataset. We will explore these two possibilities in a future work with the goal of providing classifiers based on the QCARs created from the biclusters. Anyway, these results indicate that all patients with fever and lumbar pain presented nephritis of renal pelvis origin. Practically, all the patients who did not present the disease had temperature below 37.90 Celsius degrees and absence of nausea. Likewise, a high number of patients without nausea and lumbar pain did not have the disease.

\linespread{1}
\begin{table}[]
\footnotesize
\centering
\caption{Rules for the Acute dataset.}
\label{tab:rulesAcute}
\begin{tabular}{lp{10cm}rrrr}
\toprule

\multicolumn{6}{c}{\textbf{Acute - 1st decision variable}} \\
\midrule
\# & \textbf{Rule} & \textbf{Comp.} & \textbf{Conf.} & \textbf{Lift} & \textbf{Lev.} \\
\midrule
1 & nausea\{no\}, lumbarPain\{yes\}, micturitionPain\{no\} $\Rightarrow$ 0  & 0.67 & 1.00  & 1.97  & 0.17 \\ 
2 & urinePushing\{no\}, urethraBurning\{no\} $\Rightarrow$ 0  & 0.66 & 1.00  & 1.97  & 0.16 \\ 
3 & urinePushing\{yes\}, micturitionPain\{yes\} $\Rightarrow$ 1  & 0.83 & 1.00  & 2.03  & 0.21 \\ 
4 & urinePushing\{yes\}, urethraBurning\{no\} $\Rightarrow$ 1  & 0.51 & 1.00  & 2.03  & 0.13 \\ 
\midrule

\multicolumn{6}{c}{\textbf{Acute - 2nd decision variable}} \\
\midrule
\# & \textbf{Rule} & \textbf{Comp.} & \textbf{Conf.} & \textbf{Lift} & \textbf{Lev.} \\
\midrule
1 & temperature[35.50,37.90], nausea\{no\} $\Rightarrow$ 0  & 0.98 & 1.00  & 1.71  & 0.21 \\ 
2 & nausea\{no\}, lumbarPain\{no\} $\Rightarrow$ 0  & 0.82 & 1.00  & 1.71  & 0.17 \\ 
3 & temperature[39.40,41.50], lumbarPain\{yes\} $\Rightarrow$ 1  & 0.71 & 1.00  & 2.40  & 0.20 \\ 
4 & temperature[37.90,40.30], lumbarPain\{yes\} $\Rightarrow$ 1  & 0.34 & 0.95  & 2.29  & 0.09 \\ 
\bottomrule

\end{tabular}
\end{table}
\linespread{1.5}

Table~\ref{tab:rulesCar} depicts the selected rules for the Car dataset. This dataset has 4 distinct class labels, and was the one requiring more rules to cover its instances, adding up to 54. The rules of class 2, which represents 22\% of the instances, have low completeness. So, we need many rules to cover its instances. It means that the objects (cars) classified as acceptable cars compose a heterogeneous group. Class 1 has 70\% of the instances, and Rules \#1 and \#2 have a completeness of almost 50\% for this class. Rule \#1 indicates that a car with capacity for only two persons was considered unacceptable. And rule \#2 indicates that a car with low safety rating was also considered unacceptable. So, these two attributes (carrying capacity and safety rating) are decisive to explain half of the members of class 1. Rule \#3 has also a high completeness, almost 20\%. It indicates that if a car has a very high buying price and a high or very-high maintenance price, then the car is considered unacceptable. Rules \#4 and \#5 could be joined in one, indicating that if a car has a high or very high buying price, a small luggage boot, and only a medium safety rating, then the car is considered unacceptable. The other rules for class 1 are more specific, but all of them has maximum confidence (i.e, 100\%). In fact, all the selected rules for this dataset have maximum confidence. As we have said, the rules for class 2, acceptable cars, are fragmented and most of them are very specific, having a low completeness. The main rules for this class have 6\% of completeness and they involve four attributes: buying price, maintenance price, capacity in terms of persons to carry, and estimated safety rating. The same subset of attributes, together with the size of the luggage boot, appears in the rules for the class 3 (good cars) and class 4 (very good cars). The number of instances covered by the rules presented in Table~\ref{tab:rulesCar} were 1,132, 273, 24, and 40 for class label 1, 2, 3, and 4, respectively.

\linespread{1}
\begin{table}[]
\tiny
\centering
\caption{Rules for the Car dataset.}
\label{tab:rulesCar}
\begin{tabular}{lp{10cm}rrrr}
\toprule
\# & \textbf{Rule} & \textbf{Comp.} & \textbf{Conf.} & \textbf{Lift} & \textbf{Lev.} \\
\midrule
1 & persons\{2\} $\Rightarrow$ 1  & 0.48 & 1.00  & 1.43  & 0.100 \\ 
2 & safety\{low\} $\Rightarrow$ 1  & 0.48 & 1.00  & 1.43  & 0.100 \\
3 & buying\{v-high\}, maint\{high,v-high\} $\Rightarrow$ 1  & 0.18 & 1.00  & 1.43  & 0.037 \\ 
4 & buying\{high\}, lugBoot\{small\}, safety\{med\} $\Rightarrow$ 1  & 0.04 & 1.00  & 1.43  & 0.008 \\ 
5 & buying\{v-high\}, lugBoot\{small\}, safety\{med\} $\Rightarrow$ 1  & 0.04 & 1.00  & 1.43  & 0.008 \\ 
6 & buying\{med\}, maint\{high,v-high\}, lugBoot\{small\}, safety\{med\} $\Rightarrow$ 1  & 0.02 & 1.00  & 1.43  & 0.004 \\ 
7 & buying\{high\}, doors\{2,3\}, persons\{4\}, lugBoot\{med\}, safety\{med\} $\Rightarrow$ 1  & 0.01 & 1.00  & 1.43  & 0.001 \\ 
8 & buying\{v-high\}, doors\{2,3\}, persons\{4\}, lugBoot\{med\}, safety\{med\} $\Rightarrow$ 1  & 0.01 & 1.00  & 1.43  & 0.001 \\ 
9 & buying\{med\}, maint\{high,v-high\}, persons\{4\}, safety\{high\} $\Rightarrow$ 2  & 0.06 & 1.00  & 4.50  & 0.011 \\ 
10 & buying\{high\}, maint\{low,med\}, persons\{4\}, safety\{high\} $\Rightarrow$ 2  & 0.06 & 1.00  & 4.50  & 0.011 \\ 
11 & buying\{high\}, maint\{med,high\}, persons\{4\}, safety\{high\} $\Rightarrow$ 2  & 0.06 & 1.00  & 4.50  & 0.011 \\ 
12 & buying\{v-high\}, maint\{low,med\}, persons\{4\}, safety\{high\} $\Rightarrow$ 2  & 0.06 & 1.00  & 4.50  & 0.011 \\
13 & buying\{med\}, maint\{high,v-high\}, doors\{3,4\}, persons\{more\}, safety\{high\} $\Rightarrow$ 2  & 0.03 & 1.00  & 4.50  & 0.005 \\ 
14 & buying\{med\}, maint\{high,v-high\}, doors\{4,5-more\}, persons\{more\}, safety\{high\} $\Rightarrow$ 2  & 0.03 & 1.00  & 4.50  & 0.005 \\
15 & buying\{high\}, maint\{low,med\}, doors\{3,4\}, persons\{more\}, safety\{high\} $\Rightarrow$ 2  & 0.03 & 1.00  & 4.50  & 0.005 \\ 
16 & buying\{high\}, maint\{low,med\}, doors\{4,5-more\}, persons\{more\}, safety\{high\} $\Rightarrow$ 2  & 0.03 & 1.00  & 4.50  & 0.005 \\  
17 & buying\{high\}, maint\{med,high\}, doors\{3,4\}, persons\{more\}, safety\{high\} $\Rightarrow$ 2  & 0.03 & 1.00  & 4.50  & 0.005 \\ 
18 & buying\{high\}, maint\{med,high\}, doors\{4,5-more\}, persons\{more\}, safety\{high\} $\Rightarrow$ 2  & 0.03 & 1.00  & 4.50  & 0.005 \\ 
19 & buying\{v-high\}, maint\{low,med\}, doors\{3,4\}, persons\{more\}, safety\{high\} $\Rightarrow$ 2  & 0.03 & 1.00  & 4.50  & 0.005 \\ 
20 & buying\{v-high\}, maint\{low,med\}, doors\{4,5-more\}, persons\{more\}, safety\{high\} $\Rightarrow$ 2  & 0.03 & 1.00  & 4.50  & 0.005 \\ 
21 & buying\{low\}, maint\{low,med\}, persons\{4\}, lugBoot\{small\}, safety\{med\} $\Rightarrow$ 2  & 0.02 & 1.00  & 4.50  & 0.004 \\ 
22 & buying\{low\}, maint\{med,high\}, persons\{4\}, lugBoot\{small\}, safety\{med\} $\Rightarrow$ 2  & 0.02 & 1.00  & 4.50  & 0.004 \\ 
23 & buying\{low\}, maint\{high,v-high\}, persons\{4\}, lugBoot\{small\}, safety\{high\} $\Rightarrow$ 2  & 0.02 & 1.00  & 4.50  & 0.004 \\ 
24 & buying\{low\}, maint\{high,v-high\}, persons\{4\}, lugBoot\{big\}, safety\{med\} $\Rightarrow$ 2  & 0.02 & 1.00  & 4.50  & 0.004 \\ 
25 & buying\{low\}, maint\{high,v-high\}, persons\{more\}, lugBoot\{big\}, safety\{med\} $\Rightarrow$ 2  & 0.02 & 1.00  & 4.50  & 0.004 \\ 
26 & buying\{med\}, maint\{low,med\}, persons\{4\}, lugBoot\{small\}, safety\{med\} $\Rightarrow$ 2  & 0.02 & 1.00  & 4.50  & 0.004 \\ 
27 & buying\{med\}, maint\{med,high\}, persons\{4\}, lugBoot\{small\}, safety\{high\} $\Rightarrow$ 2  & 0.02 & 1.00  & 4.50  & 0.004 \\ 
28 & buying\{med\}, maint\{med,high\}, persons\{4\}, lugBoot\{big\}, safety\{med\} $\Rightarrow$ 2  & 0.02 & 1.00  & 4.50  & 0.004 \\ 
29 & buying\{med\}, maint\{med,high\}, persons\{more\}, lugBoot\{big\}, safety\{med\} $\Rightarrow$ 2  & 0.02 & 1.00  & 4.50  & 0.004 \\ 
30 & buying\{med\}, maint\{high,v-high\}, persons\{4\}, lugBoot\{big\}, safety\{med\} $\Rightarrow$ 2  & 0.02 & 1.00  & 4.50  & 0.004 \\ 
31 & buying\{med\}, maint\{high,v-high\}, persons\{more\}, lugBoot\{med\}, safety\{high\} $\Rightarrow$ 2  & 0.02 & 1.00  & 4.50  & 0.004 \\ 
32 & buying\{med\}, maint\{high,v-high\}, persons\{more\}, lugBoot\{big\}, safety\{med\} $\Rightarrow$ 2  & 0.02 & 1.00  & 4.50  & 0.004 \\ 
33 & buying\{med\}, maint\{high,v-high\}, persons\{more\}, lugBoot\{big\}, safety\{high\} $\Rightarrow$ 2  & 0.02 & 1.00  & 4.50  & 0.004 \\ 
34 & buying\{high\}, maint\{low,med\}, persons\{4\}, lugBoot\{big\}, safety\{med\} $\Rightarrow$ 2  & 0.02 & 1.00  & 4.50  & 0.004 \\ 
35 & buying\{high\}, maint\{low,med\}, persons\{more\}, lugBoot\{med\}, safety\{high\} $\Rightarrow$ 2  & 0.02 & 1.00  & 4.50  & 0.004 \\ 
36 & buying\{high\}, maint\{low,med\}, persons\{more\}, lugBoot\{big\}, safety\{med\} $\Rightarrow$ 2  & 0.02 & 1.00  & 4.50  & 0.004 \\ 
37 & buying\{high\}, maint\{low,med\}, persons\{more\}, lugBoot\{big\}, safety\{high\} $\Rightarrow$ 2  & 0.02 & 1.00  & 4.50  & 0.004 \\ 
38 & buying\{high\}, maint\{med,high\}, persons\{4\}, lugBoot\{big\}, safety\{med\} $\Rightarrow$ 2  & 0.02 & 1.00  & 4.50  & 0.004 \\ 
39 & buying\{high\}, maint\{med,high\}, persons\{more\}, lugBoot\{med\}, safety\{high\} $\Rightarrow$ 2  & 0.02 & 1.00  & 4.50  & 0.004 \\ 
40 & buying\{high\}, maint\{med,high\}, persons\{more\}, lugBoot\{big\}, safety\{med\} $\Rightarrow$ 2  & 0.02 & 1.00  & 4.50  & 0.004 \\ 
41 & buying\{high\}, maint\{med,high\}, persons\{more\}, lugBoot\{big\}, safety\{high\} $\Rightarrow$ 2  & 0.02 & 1.00  & 4.50  & 0.004 \\ 
42 & buying\{v-high\}, maint\{low,med\}, persons\{4\}, lugBoot\{big\}, safety\{med\} $\Rightarrow$ 2  & 0.02 & 1.00  & 4.50  & 0.004 \\ 
43 & buying\{v-high\}, maint\{low,med\}, persons\{more\}, lugBoot\{med\}, safety\{high\} $\Rightarrow$ 2  & 0.02 & 1.00  & 4.50  & 0.004 \\ 
44 & buying\{v-high\}, maint\{low,med\}, persons\{more\}, lugBoot\{big\}, safety\{med\} $\Rightarrow$ 2  & 0.02 & 1.00  & 4.50  & 0.004 \\ 
45 & buying\{v-high\}, maint\{low,med\}, persons\{more\}, lugBoot\{big\}, safety\{high\} $\Rightarrow$ 2  & 0.02 & 1.00  & 4.50  & 0.004 \\ 
46 & buying\{low\}, maint\{low,med\}, persons\{4\}, lugBoot\{small\}, safety\{high\} $\Rightarrow$ 3  & 0.12 & 1.00  & 25.04  & 0.004 \\ 
47 & buying\{low\}, maint\{low,med\}, persons\{4\}, lugBoot\{big\}, safety\{med\} $\Rightarrow$ 3  & 0.12 & 1.00  & 25.04  & 0.004 \\ 
48 & buying\{low\}, maint\{low,med\}, persons\{more\}, lugBoot\{big\}, safety\{med\} $\Rightarrow$ 3  & 0.12 & 1.00  & 25.04  & 0.004 \\ 
49 & buying\{low\}, maint\{low,med\}, persons\{4\}, lugBoot\{big\}, safety\{high\} $\Rightarrow$ 4  & 0.12 & 1.00  & 26.58  & 0.004 \\ 
50 & buying\{low\}, maint\{low,med\}, persons\{more\}, lugBoot\{big\}, safety\{high\} $\Rightarrow$ 4  & 0.12 & 1.00  & 26.58  & 0.004 \\ 
51 & buying\{low\}, maint\{med,high\}, persons\{4\}, lugBoot\{big\}, safety\{high\} $\Rightarrow$ 4  & 0.12 & 1.00  & 26.58  & 0.004 \\ 
52 & buying\{low\}, maint\{med,high\}, persons\{more\}, lugBoot\{big\}, safety\{high\} $\Rightarrow$ 4  & 0.12 & 1.00  & 26.58  & 0.004 \\ 
53 & buying\{med\}, maint\{low,med\}, persons\{4\}, lugBoot\{big\}, safety\{high\} $\Rightarrow$ 4  & 0.12 & 1.00  & 26.58  & 0.004 \\ 
54 & buying\{med\}, maint\{low,med\}, persons\{more\}, lugBoot\{big\}, safety\{high\} $\Rightarrow$ 4  & 0.12 & 1.00  & 26.58  & 0.004 \\ 
\bottomrule
\end{tabular}
\end{table}
\linespread{1.5}

Table~\ref{tab:rulesHeart} presents the selected rules for the Heart dataset. We have 19 rules describing the patients with and without heart disease. Only one patient is not covered by a rule (he has heart disease). More than 30\% of the patients with heart disease presented asymptomatic chest pain, reversible defect in thal, and flat slope of the peak exercise ST segment (rule \#34). Also, more than 30\% of the patients with heart disease presented these two first characteristics together with resting electrocardiographic results having ST-T wave abnormality (showing probable or definite left ventricular hypertrophy by Estes' criteria) (rule \#33). Other 17\% of the patients with heart disease have the same result for the electrocardiographic, and are male with serum cholesterol in mg/dl in the range [274, 304] (rule \#29). A portion of 23\% of the patients with heart disease are male, with asymptomatic chest pain and one major vessel coloured by fluoroscopy (rule \#27). Among the rules of class 1 (presence of heart disease), we have six rules containing male sex and only one containing female sex. On the other hand, the rule with the highest completeness of the class 0 (rule \#8), indicates that 30\% of the patients without heart disease are female, non-exercise induced angina, and with 0 major vessels coloured by fluoroscopy. Another two rules with a high coverage of the patients without heart disease (15\%) are rules \#11 and \#13, both indicating ST depression induced by exercise relative to rest in the range [0, 0.40]. Besides this, rule \#11 indicates atypical angina chest pain and normal thal, and rule \#13 indicates resting blood pressure in the range [110, 120] and 0 major vessels coloured by fluoroscopy.

\linespread{1}
\begin{table}[]
\tiny
\centering
\caption{Rules for the Heart dataset.}
\label{tab:rulesHeart}
\begin{tabular}{lp{10cm}rrrr}
\toprule
\# & \textbf{Rule} & \textbf{Comp.} & \textbf{Conf.} & \textbf{Lift} & \textbf{Lev.} \\
\midrule
1 & age[42,46], heartRate[156,165] $\Rightarrow$ 0  & 0.04 & 1.00  & 1.80  & 0.01 \\ 
2 & age[50,54], chestPain\{non-anginal Pain\} $\Rightarrow$ 0  & 0.13 & 0.95  & 1.71  & 0.03 \\ 
3 & age[51,54], fastBSugar\{yes\}, exercIAngina\{no\} $\Rightarrow$ 0  & 0.05 & 1.00  & 1.80  & 0.01 \\ 
4 & age[53,57], heartRate[158,168] $\Rightarrow$ 0  & 0.08 & 1.00  & 1.80  & 0.02 \\ 
5 & age[54,58], bloodPres[100,110], fastBSugar\{no\}, vesselsColor\{0\} $\Rightarrow$ 0  & 0.04 & 1.00  & 1.80  & 0.01 \\ 
6 & age[55,59], fastBSugar\{no\}, heartRate[145,155], vesselsColor\{0\} $\Rightarrow$ 0  & 0.03 & 1.00  & 1.80  & 0.01 \\ 
7 & age[62,66], bloodPres[120,128], oldpeak[0,0.40] $\Rightarrow$ 0  & 0.03 & 1.00  & 1.80  & 0.01 \\ 
8 & sex\{F\}, exercIAngina\{no\}, vesselsColor\{0\} $\Rightarrow$ 0  & 0.30 & 0.96  & 1.72  & 0.07 \\ 
9 & sex\{M\}, chestPain\{non-anginal Pain\}, bloodPres[120,130], chol[226,255], electro\{normal\} $\Rightarrow$ 0  & 0.04 & 1.00  & 1.80  & 0.01 \\ 
10 & chestPain\{typical Angina\}, fastBSugar\{no\}, oldpeak[1.40,1.90] $\Rightarrow$ 0  & 0.03 & 1.00  & 1.80  & 0.01 \\ 
11 & chestPain\{atypical Angina\}, oldpeak[0,0.40], thal\{normal\} $\Rightarrow$ 0  & 0.15 & 0.96  & 1.72  & 0.03 \\ 
12 & chestPain\{non-anginal Pain\}, slope\{upsloping\}, vesselsColor\{1\} $\Rightarrow$ 0  & 0.07 & 1.00  & 1.80  & 0.02 \\ 
13 & bloodPres[110,120], oldpeak[0,0.40], vesselsColor\{0\} $\Rightarrow$ 0  & 0.15 & 0.96  & 1.72  & 0.03 \\ 
14 & bloodPres[132,142], oldpeak[0,0.50], vesselsColor\{0\} $\Rightarrow$ 0  & 0.14 & 0.95  & 1.72  & 0.03 \\ 
15 & bloodPres[150,160], fastBSugar\{yes\} $\Rightarrow$ 0  & 0.04 & 1.00  & 1.80  & 0.01 \\ 
16 & chol[204,234], electro\{LVH\}, vesselsColor\{0\} $\Rightarrow$ 0  & 0.11 & 1.00  & 1.80  & 0.03 \\ 
17 & chol[295,325], electro\{normal\}, exercIAngina\{no\} $\Rightarrow$ 0  & 0.08 & 1.00  & 1.80  & 0.02 \\ 
18 & electro\{normal\}, slope\{upsloping\}, vesselsColor\{0\}, thal\{normal\} $\Rightarrow$ 0  & 0.25 & 0.95  & 1.71  & 0.06 \\ 
19 & oldpeak[0.30,0.80], thal\{normal\} $\Rightarrow$ 0  & 0.15 & 0.96  & 1.72  & 0.03 \\ 
20 & age[48,52], sex\{M\}, fastBSugar\{no\}, exercIAngina\{no\}, thal\{reversable Defect\} $\Rightarrow$ 1  & 0.05 & 1.00  & 2.25  & 0.01 \\ 
21 & age[57,60], bloodPres[150,160], fastBSugar\{no\}, electro\{LVH\} $\Rightarrow$ 1  & 0.04 & 1.00  & 2.25  & 0.01 \\ 
22 & age[58,62], vesselsColor\{2\} $\Rightarrow$ 1  & 0.10 & 1.00  & 2.25  & 0.02 \\ 
23 & age[59,63], sex\{F\}, electro\{normal\}, slope\{flat\} $\Rightarrow$ 1  & 0.04 & 1.00  & 2.25  & 0.01 \\ 
24 & age[65,67], chestPain\{asymptomatic\}, oldpeak[0.60,1] $\Rightarrow$ 1  & 0.04 & 1.00  & 2.25  & 0.01 \\ 
25 & age[66,70], exercIAngina\{yes\} $\Rightarrow$ 1  & 0.07 & 1.00  & 2.25  & 0.02 \\ 
26 & sex\{M\}, chestPain\{non-anginal Pain\}, slope\{flat\}, vesselsColor\{1\} $\Rightarrow$ 1  & 0.05 & 1.00  & 2.25  & 0.01 \\ 
27 & sex\{M\}, chestPain\{asymptomatic\}, vesselsColor\{1\} $\Rightarrow$ 1  & 0.23 & 0.97  & 2.17  & 0.06 \\ 
28 & sex\{M\}, bloodPres[136,146], fastBSugar\{no\}, oldpeak[1.60,2] $\Rightarrow$ 1  & 0.06 & 1.00  & 2.25  & 0.01 \\ 
29 & sex\{M\}, chol[274,304], electro\{LVH\} $\Rightarrow$ 1  & 0.17 & 0.95  & 2.15  & 0.04 \\ 
30 & sex\{M\}, heartRate[124,132], oldpeak[0.80,1.20] $\Rightarrow$ 1  & 0.04 & 1.00  & 2.25  & 0.01 \\ 
31 & chestPain\{asymptomatic\}, bloodPres[130,140], heartRate[103,111] $\Rightarrow$ 1  & 0.04 & 1.00  & 2.25  & 0.01 \\ 
32 & chestPain\{asymptomatic\}, bloodPres[150,160], fastBSugar\{no\}, thal\{reversable Defect\} $\Rightarrow$ 1  & 0.07 & 1.00  & 2.25  & 0.02 \\ 
33 & chestPain\{asymptomatic\}, electro\{LVH\}, thal\{reversable Defect\} $\Rightarrow$ 1  & 0.32 & 0.97  & 2.19  & 0.08 \\ 
34 & chestPain\{asymptomatic\}, slope\{flat\}, thal\{reversable Defect\} $\Rightarrow$ 1  & 0.33 & 0.95  & 2.14  & 0.08 \\ 
35 & bloodPres[124,132], heartRate[131,141], thal\{reversable Defect\} $\Rightarrow$ 1  & 0.07 & 1.00  & 2.25  & 0.02 \\ 
36 & bloodPres[130,140], chol[330,353] $\Rightarrow$ 1  & 0.04 & 1.00  & 2.25  & 0.01 \\ 
37 & bloodPres[132,140], oldpeak[2.60,3.10] $\Rightarrow$ 1  & 0.04 & 1.00  & 2.25  & 0.01 \\ 
38 & fastBSugar\{no\}, oldpeak[3.40,3.80], slope\{flat\} $\Rightarrow$ 1  & 0.04 & 1.00  & 2.25  & 0.01 \\ \bottomrule
\end{tabular}
\end{table}
\linespread{1.5}

The selected rules for the Voting dataset are exhibited in Table~\ref{tab:rulesVoting}. It is possible to verify that 5 rules were used to describe the Republicans and 8 rules were used to describe the Democrats. Almost all Democrats, 92\%, are identified by the negative vote in the physician-fee-freeze (rule \#6). At the same time, 83\% of the Republicans are identified by the positive vote in the physician-fee-freeze and by the negative vote in the adoption-of-the-budget-resolution (rule \#1). When we have the same attributes in both rules of class 0 or class 1, they always appear with the opposite values, for instance, physician-fee-freeze, education-spending, and adoption-of-the-budget-resolution. Rule \#10 shows that the Democrats voted in favour of the adoption-of-the-budget-resolution, and synfuels-corporation-cutback. Whereas, rule \#2 shows that the Republicans voted against these two topics, and also against duty-free-exports. Similarly, rule \#11 shows that Democrats voted in favour of synfuels-corporation-cutback and against education-spending. Rule \# 4 indicates that Republicans voted against synfuels-corporation-cutback, in favour of education-spending, and against anti-satellite-test-ban. From the 4 voters not covered by any rule of Table~\ref{tab:rulesVoting}, 3 of them have too many missing values. The fourth voter is a Democrat exhibiting an abnormal pattern.

Notice that the Voting dataset contains only binary attributes, thus allowing a direct comparison with results provided by traditional FPM/FCA algorithms. Our rules contain positive (yes) and negative (no) responses for the attributes. Rules mined in a traditional way by FPM/FCA algorithms contain only positive answers. Of course, it is possible to use strategies, such as the itemization, that creates an augmented binary matrix with twice the columns of the original one. Depending on the number of attributes, this matrix augmentation may become computationally prohibitive.

\linespread{1}
\begin{table}[]
\footnotesize
\centering
\caption{Rules for the Voting dataset.}
\label{tab:rulesVoting}
\begin{tabular}{lp{10cm}rrrr}
\toprule
\# & \textbf{Rule} & \textbf{Comp.} & \textbf{Conf.} & \textbf{Lift} & \textbf{Lev.} \\
\midrule
1 & budgetRes\{no\}, physicianFF\{yes\} $\Rightarrow$ 0  & 0.83 & 0.96  & 2.48  & 0.19 \\ 
2 & budgetRes\{no\}, sfCorpCut\{no\}, dutyFree\{no\} $\Rightarrow$ 0  & 0.62 & 0.97  & 2.52  & 0.15 \\ 
3 & physicianFF\{yes\}, admSA\{yes\} $\Rightarrow$ 0  & 0.56 & 0.96  & 2.48  & 0.13 \\ 
4 & antiSatelliteTT\{no\}, sfCorpCut\{no\}, eduSpending\{yes\} $\Rightarrow$ 0  & 0.55 & 0.97  & 2.51  & 0.13 \\ 
5 & ES-aid\{yes\}, antiSatelliteTT\{yes\}, mxMissile\{no\}, sfCorpCut\{no\} $\Rightarrow$ 0  & 0.11 & 0.95  & 2.46  & 0.03 \\ 
6 & physicianFF\{no\} $\Rightarrow$ 1  & 0.92 & 0.99  & 1.62  & 0.21 \\ 
7 & aidNicaraguaC\{yes\}, eduSpending\{no\} $\Rightarrow$ 1  & 0.70 & 0.96  & 1.56  & 0.15 \\ 
8 & hInfants\{yes\}, mxMissile\{yes\} $\Rightarrow$ 1  & 0.43 & 0.96  & 1.56  & 0.10 \\ 
9 & budgetRes\{yes\}, immigration\{no\} $\Rightarrow$ 1  & 0.43 & 0.96  & 1.56  & 0.10 \\ 
10 & budgetRes\{yes\}, sfCorpCut\{yes\} $\Rightarrow$ 1  & 0.40 & 0.97  & 1.58  & 0.09 \\ 
11 & sfCorpCut\{yes\}, eduSpending\{no\} $\Rightarrow$ 1  & 0.36 & 0.97  & 1.58  & 0.08 \\ 
12 & wProject\{yes\}, superfundRS\{no\} $\Rightarrow$ 1  & 0.24 & 0.98  & 1.60  & 0.06 \\ 
13 & wProject\{yes\}, dutyFree\{yes\} $\Rightarrow$ 1  & 0.24 & 0.97  & 1.58  & 0.05 \\ 
\bottomrule
\end{tabular}
\end{table}
\linespread{1.5}

Finally, Table~\ref{tab:rulesZoo} is devoted to present the selected rules for the Zoo dataset. Only one rule is necessary to describe the animals of classes 1, 2, 4, 5, and 6. Classes 3 and 7 are each one described by two rules. A peculiar aspect of the obtained results is that we have some attributes appearing in all rules, such as milk and feathers. Milk has the value yes only for class 1, and feathers has the value yes only for class 2. In fact, these attributes alone can fully describe these classes. They appear together with other attributes because the biclusters are maximal, and all discriminant aspects present in the instances of a class are shown. For instance, all animals from class 1 do not have feathers, are mammals, have backbone, breath, and are not venomous.

\linespread{1}
\begin{table}[]
\tiny
\centering
\caption{Rules for the Zoo dataset.}
\label{tab:rulesZoo}
\begin{tabular}{lp{10cm}rrrr}
\toprule
\# & \textbf{Rule} & \textbf{Comp.} & \textbf{Conf.} & \textbf{Lift} & \textbf{Lev.} \\
\midrule
1 & feathers\{no\}, milk\{yes\}, backbone\{yes\}, breathes\{yes\}, venomous\{no\} $\Rightarrow$ 1  & 1.00 & 1.00  & 2.46  & 0.24 \\ 
2 & hair\{no\}, feathers\{yes\}, eggs\{yes\}, milk\{no\}, toothed\{no\}, backbone\{yes\}, breathes\{yes\}, venomous\{no\}, fins\{no\}, legs\{2\}, tail\{yes\} $\Rightarrow$ 2  & 1.00 & 1.00  & 5.05  & 0.16 \\ 
3 & hair\{no\}, feathers\{no\}, eggs\{yes\}, milk\{no\}, airborne\{no\}, aquatic\{no\}, backbone\{yes\}, breathes\{yes\}, fins\{no\}, tail\{yes\}, domestic\{no\} $\Rightarrow$ 3  & 0.80 & 1.00  & 20.20  & 0.04 \\ 
4 & hair\{no\}, feathers\{no\}, milk\{no\}, airborne\{no\}, predator\{yes\}, toothed\{yes\}, backbone\{yes\}, fins\{no\}, legs\{0\}, tail\{yes\}, domestic\{no\}, catsize\{no\} $\Rightarrow$ 3  & 0.60 & 1.00  & 20.20  & 0.03 \\ 
5 & hair\{no\}, feathers\{no\}, eggs\{yes\}, milk\{no\}, airborne\{no\}, aquatic\{yes\}, toothed\{yes\}, backbone\{yes\}, breathes\{no\}, fins\{yes\}, legs\{0\}, tail\{yes\} $\Rightarrow$ 4  & 1.00 & 1.00  & 7.77  & 0.11 \\ 
6 & hair\{no\}, feathers\{no\}, eggs\{yes\}, milk\{no\}, airborne\{no\}, aquatic\{yes\}, toothed\{yes\}, backbone\{yes\}, breathes\{yes\}, fins\{no\}, legs\{4\}, domestic\{no\}, catsize\{no\} $\Rightarrow$ 5  & 1.00 & 1.00  & 25.25  & 0.04 \\ 
7 & feathers\{no\}, eggs\{yes\}, milk\{no\}, aquatic\{no\}, toothed\{no\}, backbone\{no\}, breathes\{yes\}, fins\{no\}, legs\{6\}, tail\{no\}, catsize\{no\} $\Rightarrow$ 6  & 1.00 & 1.00  & 12.62  & 0.07 \\ 
8 & hair\{no\}, feathers\{no\}, eggs\{yes\}, milk\{no\}, airborne\{no\}, toothed\{no\}, backbone\{no\}, fins\{no\}, legs\{0\}, tail\{no\}, domestic\{no\}, catsize\{no\} $\Rightarrow$ 7  & 0.40 & 1.00  & 10.10  & 0.04 \\ 
9 & hair\{no\}, feathers\{no\}, milk\{no\}, airborne\{no\}, predator\{yes\}, toothed\{no\}, backbone\{no\}, fins\{no\}, domestic\{no\} $\Rightarrow$ 7  & 0.80 & 1.00  & 10.10  & 0.07 \\ 
\bottomrule
\end{tabular}
\end{table}
\linespread{1.5}

\section{Concluding remarks}
\label{sec:conclusion}

In this paper, we provided an enumerative biclustering algorithm to mine all maximal biclusters directly in mixed-attribute datasets, with or without missing values. A mixed-attribute dataset may be represented by a mixed-data matrix having any kind of attribute in each column, ranging from numerical (discrete or continuous) to categorical (ordinal or nominal). Of course, when a single type of data is present, such as all attributes being binary, or all attributes being real values, our proposal will work as well. As far as we know, all the alternative biclustering algorithms in the literature also devoted to handle mixed-attribute datasets must rely on discretization and/or itemization routines, thus involving information loss. This new algorithm is an extension of an existing proposal to mine constant values on columns (CVC) numerical biclusters, denoted RIn-Close\_CVC \cite{VeronezeEtAl2017}, and keeps the four key properties of its predecessor: efficiency, completeness, correctness, and non-redundancy. The extension does not require additional computational cost, so that the extension exhibits the same worst-case time complexity of the original algorithm.

Additionally, the strong connection between biclustering and frequent pattern mining (FPM) is widely explored to (1) present the biclusters in a user-friendly and intuitive form, by automatically converting them to quantitative class association rules (QCARs), and (2) select a subset of meaningful biclusters from the enumerative solution by threshold indices derived from consolidated FPM metrics, more specifically confidence and lift. Moreover, our experimental results indicated that the QCARs extracted from the biclusters are valuable and automatic means of providing useful and relevant interpretable models of a dataset.

In addition to the selection of biclusters based on FPM metrics, we also provided a simple heuristic to select a small but still representative group of biclusters. Our results showed that these biclusters yield a parsimonious set of relevant rules for discriminating the class labels.

In a future work, the interplay between RIn-Close\_CVC biclustering and QCARs will be further explored in the context of associative classification, which is an emerging FPM research field devoted to the synthesis of high-performance rule-based classifiers \cite{LiuEtAl1998, NguyenEtAl2015}. There are open issues in the selection of the mined CARs when building rule-based classifiers. Our intention is to address those open issues and to incorporate QCARs and fuzzy CARs \cite{AntonelliEtAl2015}.

Given that the enumerated biclusters are maximal, the associated QCARs are not the most parsimonious ones, because we are mainly focused on representative power. As another future perspective of the research, we are going to explore strategies for selecting a small set of the most informative attributes to discriminate between class labels.

\section*{Acknowledgments}

F. J. Von Zuben would like to thank CNPq (process 309115/2014-0) for the financial support.


\linespread{1}

{\footnotesize
\bibliography{tese}}

\vfill
{\footnotesize
\textbf{Rosana Veroneze} is PhD in Computer Engineering at the University of Campinas (Unicamp). Her research interests include data mining, machine learning, computational intelligence, and bioinspired computing.

\textbf{Fernando J. Von Zuben} is a Full Professor at the Department of Computer Engineering and Industrial Automation, School of Electrical and Computer Engineering, University of Campinas (Unicamp). The main topics of his research are computational intelligence, bioinspired computing, multivariate data analysis, and machine learning.
}

\end{document}